\documentclass[aps,prl,reprint,twocolumn,superscriptaddress]{revtex4-1}

\usepackage{euscript}
\usepackage{amssymb}
\usepackage{amsfonts}
\usepackage{amsbsy}
\usepackage{epsfig}
\usepackage{amsthm}
\usepackage{amscd}
\usepackage{amstext}
\usepackage{verbatim}
\usepackage{amsmath}
\usepackage{cancel}
\usepackage{capt-of}
\usepackage{empheq}
\usepackage{subfigure}
\usepackage{xcolor}

\usepackage[pdftex]{hyperref}
\hypersetup{ 
colorlinks=true, 
linkcolor=black, 
citecolor=red, 
}

\usepackage{amssymb}

\def\ben{\begin{equation}}
\def\een{\end{equation}}
\def\half{{\textstyle{\frac{1}{2}}}}
\def\qtr{{\textstyle{\frac{1}{4}}}}
\let\a=\alpha    
   \let\k=\kappa
     
\let\s=\sigma \let\t=\tau

\let\pa=\partial
\def\be{\begin{equation}}
\def\ee{\end{equation}}
\def\beq{\begin{equation}}
\def\eeq{\end{equation}}
\def\ba{\begin{array}}
\def\ea{\end{array}}

\def\dalemb#1#2{{\vbox{\hrule height .#2pt
       \hbox{\vrule width.#2pt height#1pt \kern#1pt
               \vrule width.#2pt}
       \hrule height.#2pt}}}

\newcommand{\bea}{\begin{eqnarray}}
\newcommand{\eea}{\end{eqnarray}}

\newcommand{\Tr}{{\rm Tr} }

\makeatletter
\newcommand*\bigcdot{\mathpalette\bigcdot@{.5}}
\newcommand*\bigcdot@[2]{\mathbin{\vcenter{\hbox{\scalebox{#2}{$\m@th#1\bullet$}}}}}
\makeatother

\renewcommand{\eqref}[1]{(\ref{#1})}

\def\ocal{{\mathcal{O}}}

\begin{document}
\frenchspacing

\title{Bad metallic transport in a modified Hubbard model}
\author{Connie H. Mousatov, Ilya Esterlis, Sean A. Hartnoll\vspace{0.4cm} \\ {\it Department of Physics, Stanford University,}\\
{\it Stanford, California, USA}}



\begin{abstract}
Strongly correlated metals often display anomalous transport, including $T$-linear resistivity above the Mott-Ioffe-Regel limit. We introduce a tractable microscopic model for bad metals, by supplementing the well-known Hubbard model --- with hopping $t$ and on-site repulsion $U$ --- with a `screened Coulomb' interaction between charge densities that decays exponentially with spatial separation. This interaction entirely lifts the extensive degeneracy in the spectrum of the $t=0$ Hubbard model, allowing us to fully characterize the small $t$ electric, thermal and thermoelectric transport in our strongly correlated model.
Throughout the phase diagram we observe $T$-linear resistivity above the Mott-Ioffe-Regel limit, together with strong violation of the Weidemann-Franz law and a large thermopower that can undergo sign change.


\end{abstract} 

\maketitle

{\it Introduction.---} In conventional metals, electrical resistance arises from the microscopic scattering of electronic quasiparticles. This paradigm is challenged in bad metals, where the resistivity grows with temperature above the Mott-Ioffe-Regel (MIR) limit \cite{PhysRevLett.74.3253}. Such behavior is widely observed in strongly correlated materials at high temperatures \cite{RevModPhys.75.1085,hussey}, and hints at non-quasiparticle transport which must be understood along radically different lines than traditional Boltzmann theory.

High temperature, bad metallic regimes of strongly correlated materials are often far from the battleground of multiple low temperature competing orders. Indeed, bad metals exhibit similarities across many materials, including an often noted $T$-linear resistivity \cite{Hartnoll:2014lpa}. Despite suggestive universal behavior, the understanding of bad metals has been hampered by the lack of a microscopic, theoretical model in which the resistivity can be computed without artificial control parameters. To this end, we introduce a realistic modification of the widely-studied Hubbard model for correlated electrons that allows us to obtain explicit results for high temperature, non-quasiparticle, bad metal transport.

{\it The model.---} We will study the lattice Hamiltonian
\be\label{eq:model}
H = t \sum_{\langle ij\rangle,s}  c^\dagger_{i s} c^{\phantom{\dagger}}_{j s}  + U \sum_i n_{i\uparrow}n_{i\downarrow}  + \frac{V}{2} \sum_{i \neq j}e^{ - |\vec x_i - \vec x_j|/\ell} \, n_i n_j \,.
\ee
As usual the density $n_i = \sum_s c^\dagger_{i s} c^{\phantom{\dagger}}_{i s}$, with $s \in \{\uparrow,\downarrow\}$ the fermion spin. The positions $\vec x_i = a \vec \imath$ form a two dimensional square lattice.
The first two terms in (\ref{eq:model}) comprise the usual Hubbard model, with hopping $t$ over nearest neighbours $\langle ij \rangle$ and on-site repulsion $U$. The final `screened Coulomb' interaction is short range, but not strictly finite range. This last term differentiates the model from the Hubbard model (which has $V=0$), and also from finite range extensions thereof, and is essential for our results. In particular, this modification allows us to obtain explicit and finite results for transport coefficients in the weak hopping regime $t \ll \{ k_BT, U, V \}$. Here $T$ is the temperature. These temperatures are higher than those of observed bad metals; they pertain instead to recent transport experiments in cold atomic gases \cite{Bakr}. Our immediate objective is rather to obtain controlled and physically transparent bad metal transport.


Small $t$ transport in the Hubbard model has been studied in a number of works  \cite{PhysRevB.10.2186, doi:10.1063/1.2712775}. However, the spectrum of the Hubbard model with $t=0$ is extremely degenerate, with excitations occupying either the single-site upper or lower Hubbard band. In contrast, 
the new interaction in the model (\ref{eq:model}) --- that is exponentially localized to within a microscopic range $\ell$ but not strictly finite range ---  is sufficient to split the extensive degeneracy of the $t=0$ theory. This allows us to use conventional non-degenerate perturbation theory in small $t$ to obtain a low energy spectral density, and hence transport coefficients, that are finite in the infinite volume limit.

All of the terms in the $U$ and $V$ interactions in (\ref{eq:model}) commute. This means that all computations in small $t$ perturbation theory can be evaluated using {\it classical} Monte Carlo simulations in the $t=0$ theory. This statistical description of bad metal transport is an immense simplification. The statistical regime is intrinsically incoherent and distinct from Boltzmann-Drude theory, as emphasized in \cite{PhysRevB.73.035113}. Using classical Monte Carlo, we are able to work with a large system size in two dimensions, and furthermore study the entire filling range $0 \leq n \leq 2$ and obtain the full thermoelectric conductivity matrix.

{\it The conductivity.---} To leading order at small hopping $t$, the conductivity is computed as follows. At $t=0$, occupation number configurations $\{n\}$ define eigenstates of charge $N_{\{n\}} = e \sum_{is} n_{is}$ and energy $E_{\{n\}} = \frac{1}{2} \sum_{is} n_{is} \epsilon_{is}$, with on-site energies $\epsilon_{i s} = U n_{i\bar s} + V \sum_{j \neq i} e^{ - |\vec x_i - \vec x_j|/\ell} \, n_j$. Here
$n_{i \bar s}$ is the number of electrons at site $i$ with opposite spin to $s$. Using classical Monte Carlo simulation, typical configurations $\{n\}$ are generated for a given temperature and filling. We give technical details in the Supplementary Material. The real and dissipative electrical conductivity is a weighted sum over these configurations
\be\label{eq:sigmamain}
\sigma_1(\omega) = \frac{2 e^2}{h} \frac{(\pi a t)^2}{\hbar \, \text{vol}} f(\omega) \sum_{\{n\}} \frac{e^{-\beta (E_{\{n\}}-\mu N_{\{n\}})}}{\mathcal{Z}} \sum_{i,s} \Delta_{is}(\omega) \,.
\ee
Here vol is the volume and $f(\omega) = (1 - e^{-\beta \hbar \omega})/\hbar \omega$. The inverse temperature $\beta \equiv 1/(k_B T)$ and the partition function $\mathcal{Z} = \sum_{\{n\}} e^{-\beta (E_{\{n\}}-\mu N_{\{n\}})}$. Given a configuration, the spectral weight $\Delta_{is}(\omega)$ counts the number of excitations with energy $\hbar \omega$ that can be generated with a single hop between neighbouring sites. $\Delta_{is}(\omega)$ has units of inverse frequency and is defined precisely in the Supplementary Material. Analogous formulae exist for the thermoelectric conductivity $\a$ and the thermal conductivity $\kappa$, and are also given in the Supplementary Material.

The expression (\ref{eq:sigmamain}) is strictly only valid for $\hbar \omega \gtrsim t$. At lower frequencies non-perturbative localization physics could potentially deplete the density of states $\Delta_{is}(\omega)$. This concern is addressed in the discussion section below. We proceed to use (\ref{eq:sigmamain}) to obtain dc transport observables.

\begin{figure}[h]
\hspace{0.8cm}\includegraphics[width=1.45in]{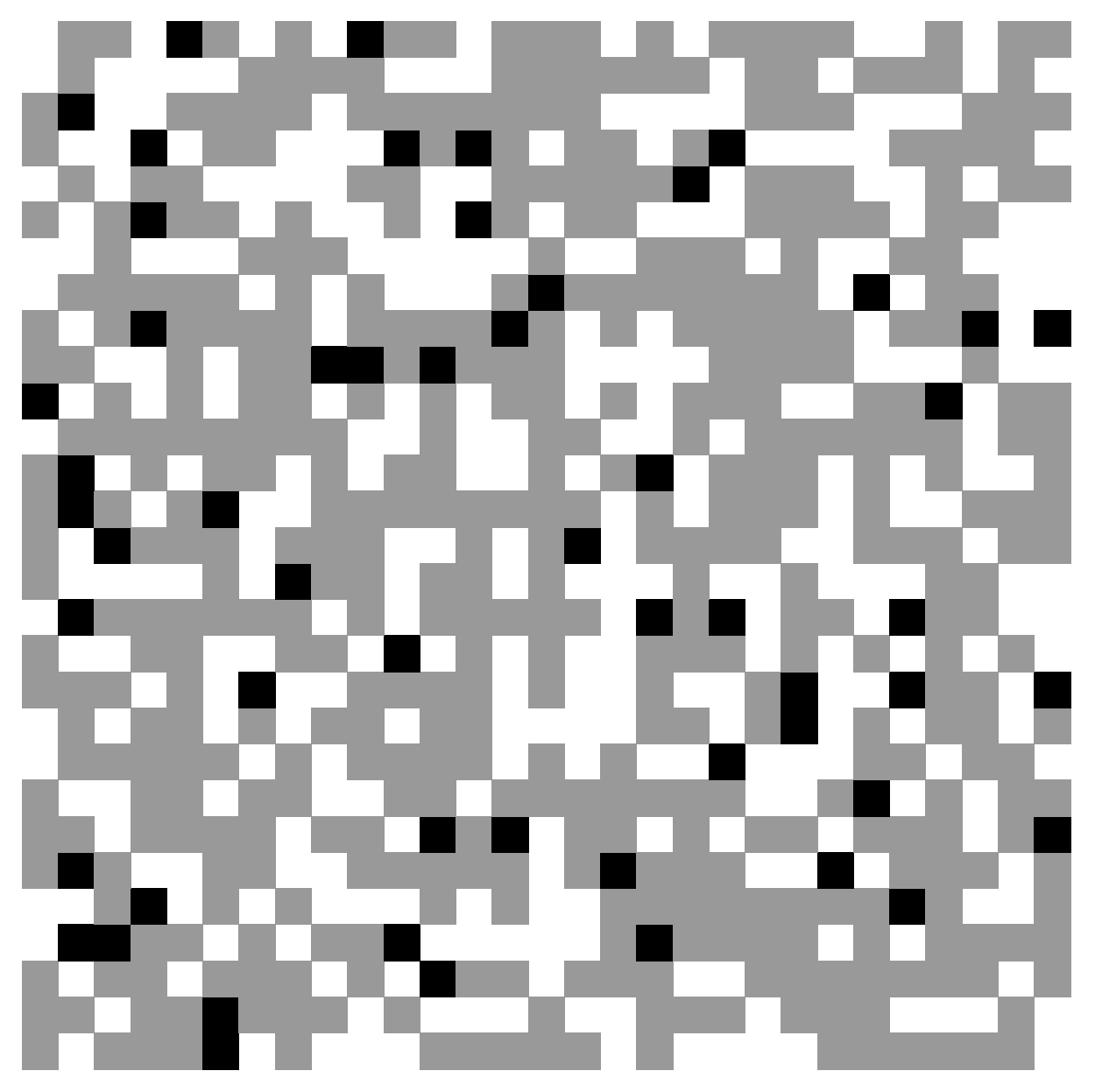}
\includegraphics[width=1.45in]{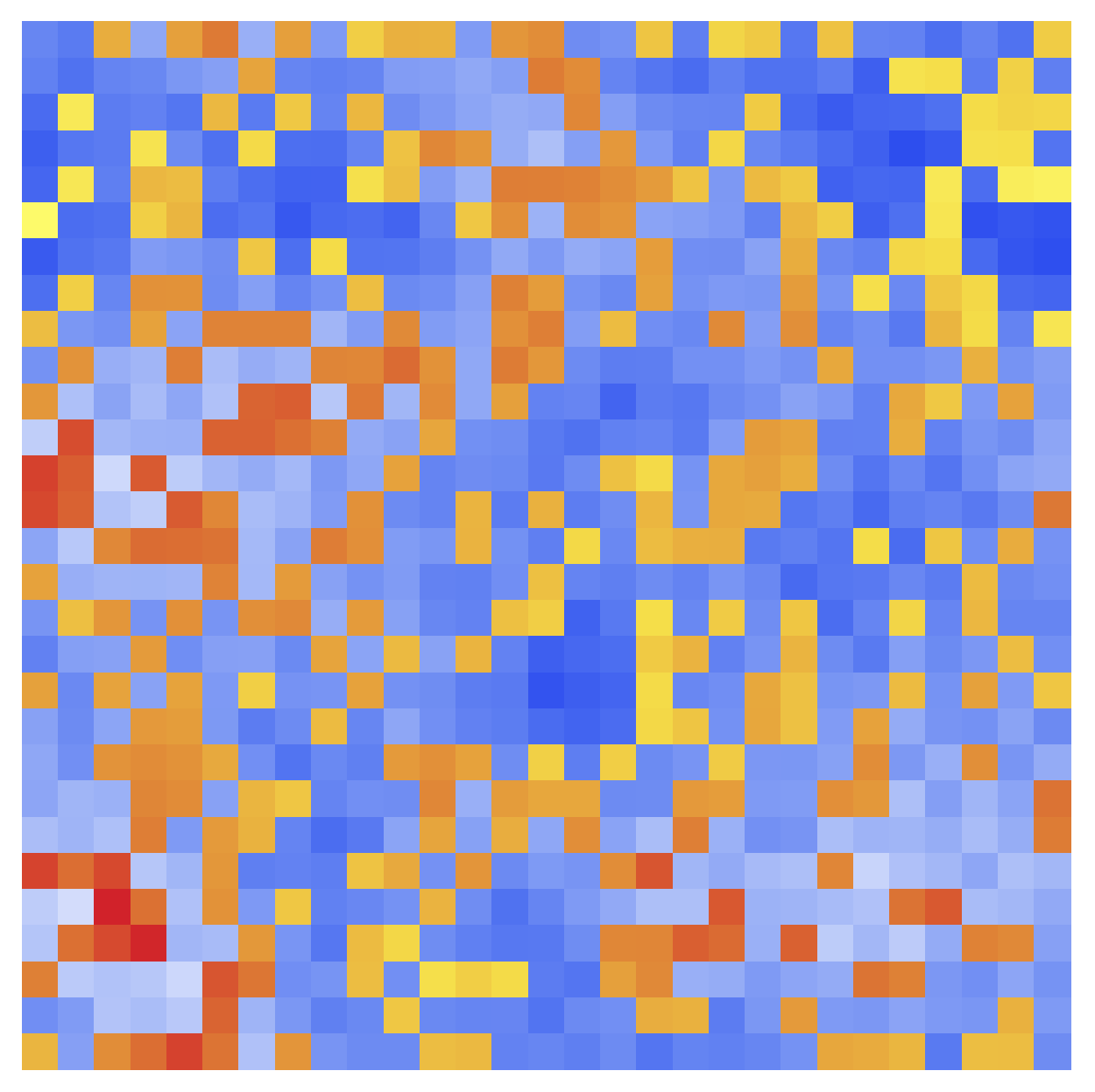} \\
\includegraphics[width=\columnwidth]{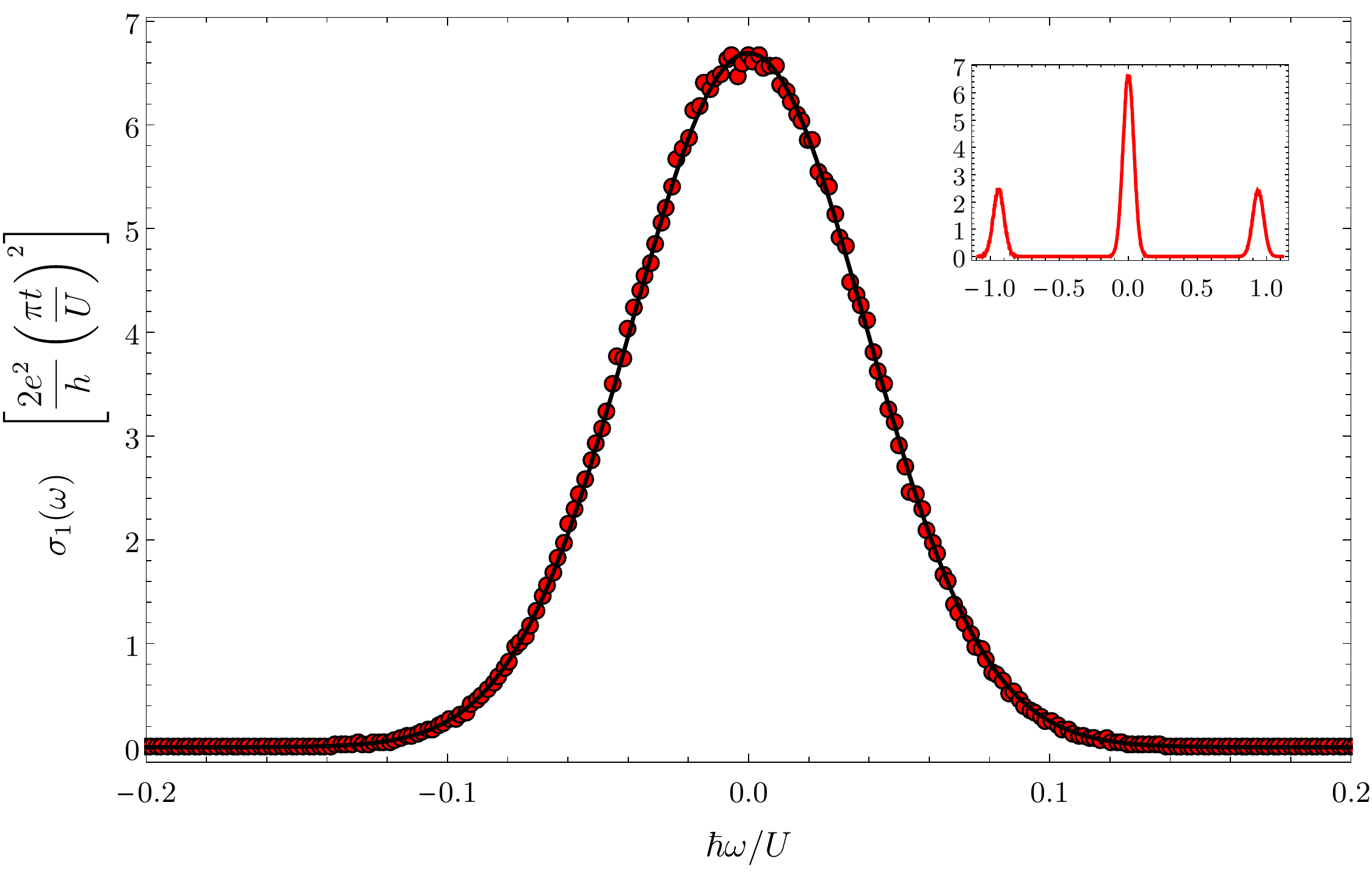}
\caption{On-site energies for a typical configuration with a $29 \times 29$ lattice at temperature $k_B T = 0.76 U$, coupling $V = 0.1 U$, range $\ell = 2 a$ and filling $n = 0.63$. Top left shows the occupation numbers for the configuration (white is unoccupied, gray is singly-occupied and black is doubly-occupied). Top right shows the on-site potentials for up spins, $\epsilon_{i\uparrow}$, generated by this configuration. A broadened upper (red/yellow) and lower (blue) Hubbard band are seen. Bottom shows the corresponding low frequency conductance peak. The solid line shows a fit to a Gaussian. The inset shows the optical conductivity over a wider frequency range, including transitions between the lower and upper Hubbard bands.}
\label{fig:Histogram}
\end{figure}

Figure \ref{fig:Histogram} shows a representative occupation number configuration, together with the corresponding on-site energies $\epsilon_{i\uparrow}$. Differences of neighboring on-site energies determine $\Delta_{is}(\omega)$ and hence the optical conductivity, also shown in the figure. The conductivity is computed using 15000 weighted configurations in (\ref{eq:sigmamain}).
The optical conductivity displays transitions between lower and upper `Hubbard bands' together with a low frequency conductance peak.
In the $t=0$ Hubbard model, the optical conductivity is a sum of delta functions at $\omega  =0, \pm U$. In Figure \ref{fig:Histogram} these peaks have been broadened, leading to a finite dc conductivity. This occurs because the exponentially localized interaction $V$, with any range $\ell > \ell_\star \approx 1.76 a$, lifts the extensive degeneracy of the $t=0$ Hubbard model, as we show in the Supplementary Material.

The low frequency peak in figure \ref{fig:Histogram} is Gaussian, in contrast to a conventional Lorentzian Drude peak. A Gaussian peak is also seen in the high temperature expansion of a hard boson model \cite{PhysRevB.81.054512}, and indicates that the energy differences contributing to $\sigma(\omega)$ are essentially random.

{\it Transport results.---} We will work throughout with the values $V =0.1U$ and $\ell = 2a$. Thus the exponential interaction is microscopically short range and the small value of $V$ means that results can be compared meaningfully  to the Hubbard model. The hopping $t \ll \{U,V, k_B T \}$.

\begin{figure}[h]
\centering
\includegraphics[width=\columnwidth]{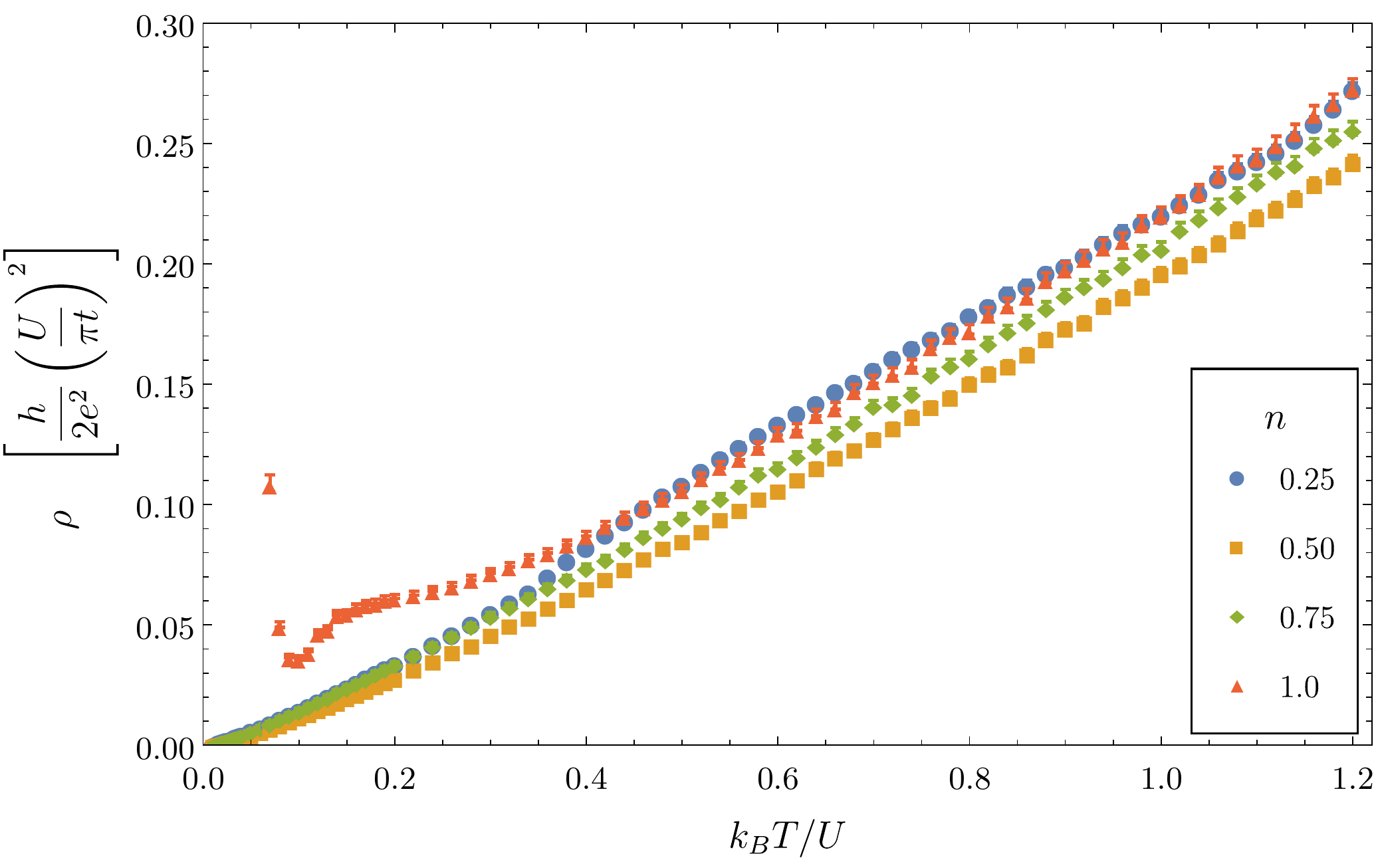}
\caption{Resistivity as a function of temperature. Statistical errors are shown.}
\label{fig:rhoTlow}
\end{figure}

The resistivity for $t \ll k_B T \lesssim U$ is shown for various fillings in figure \ref{fig:rhoTlow}. Away from the Mott insulating upturn at $n=1$, the resistivity is approximately $T$-linear, with some weak curvature at lower temperatures. The magnitude of the resistivity is $\rho \sim h/e^2 \times U^2/t^2 \gg h/e^2$ throughout, so the system is a bad metal.

\begin{figure}[h]
\centering
\includegraphics[width=0.5\columnwidth]{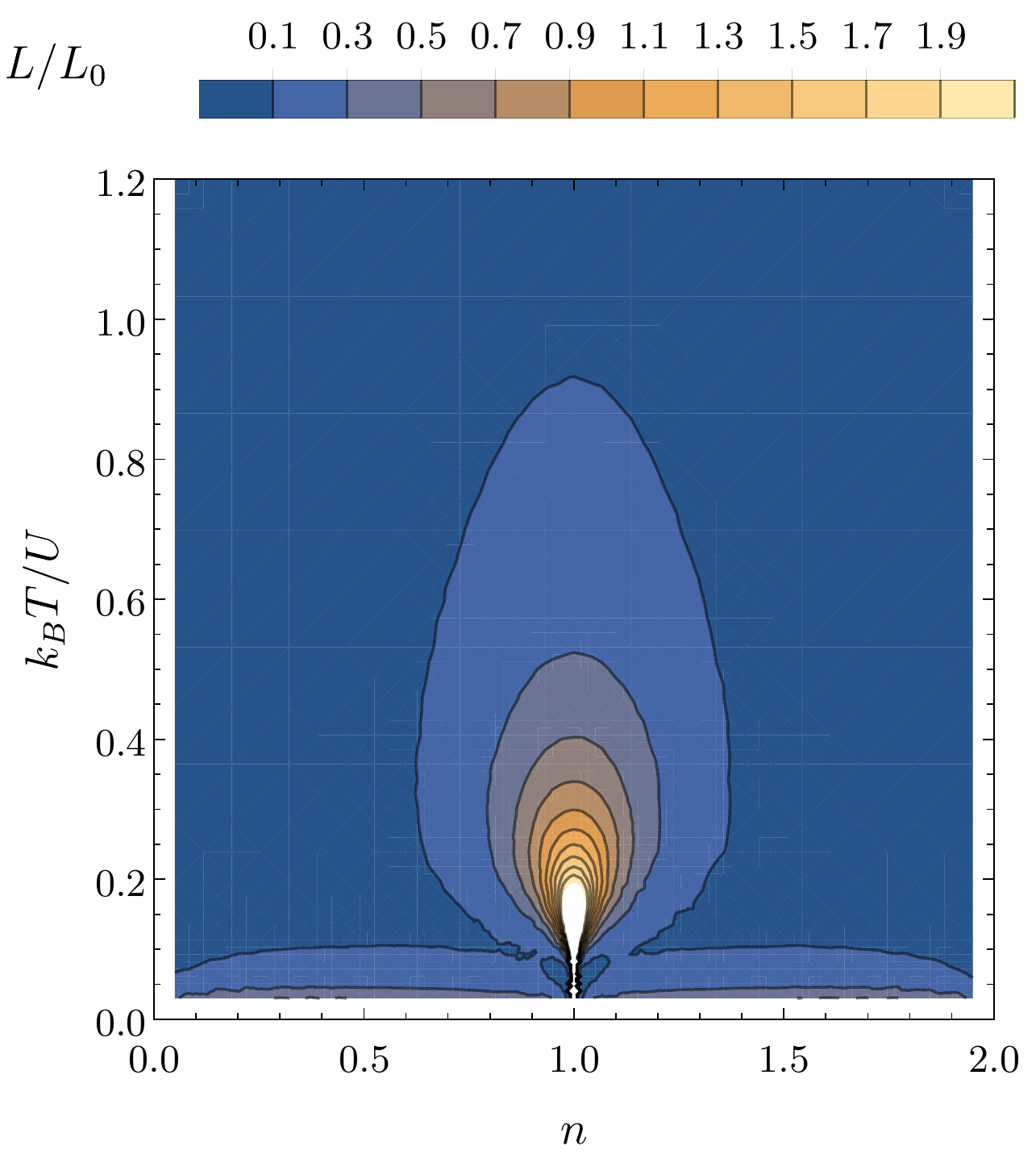}\includegraphics[width=0.5\columnwidth]{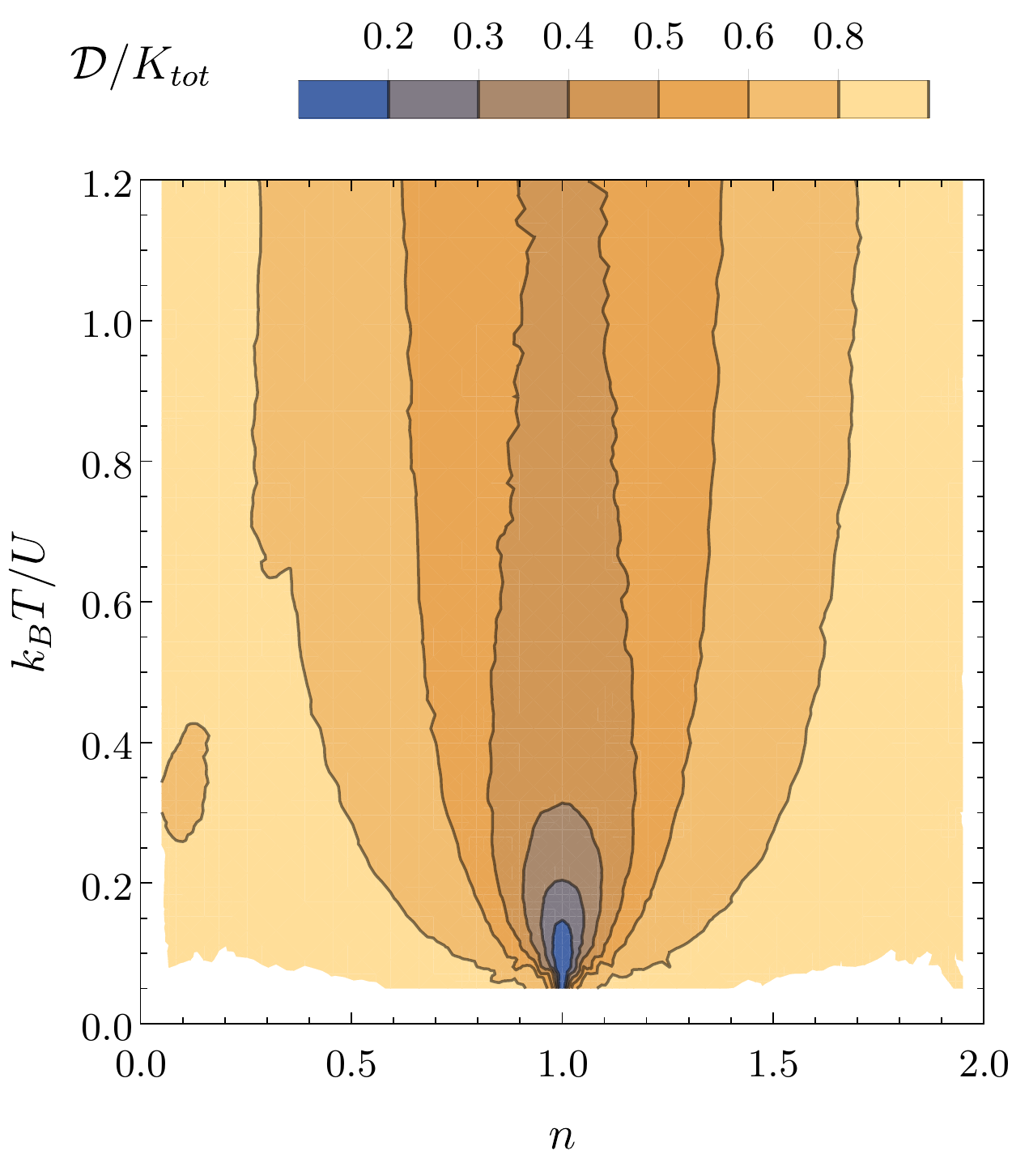}
\caption{Left: Violation of the Weidemann-Franz law across the phase diagram. The Sommerfeld value $L_0 \equiv \pi^2/3 \times (k_B/e)^2$. Right: Fraction of the electronic kinetic energy in the conductance peak --- as measured by ${\mathcal D}/K_\text{tot}$.}
\label{fig:WFTlow}
\end{figure}

Themoelectric and thermal transport are usefully quantified by the thermopower $S \equiv \alpha/\sigma$ and Lorenz ratio $L \equiv \kappa/(\sigma T)$, respectively. Figure \ref{fig:WFTlow} shows the Lorenz ratio for $t \ll k_B T \lesssim U$. Strong violation of the Weidemann-Franz (WF) law is seen across the entire phase diagram: $L \ll L_0$, the Sommerfeld value, almost everywhere, except for just above the Mott regime, where $L \gg L_0$. The WF law is not expected to hold at these high temperatures, but $L_0$ remains a useful yardstick for the relative efficacy of thermal and charge transport.
For example, $L \ll L_0$ has been observed recently in the anomalous bad metal phase of VO$_2$ \cite{Lee371}. This is a strongly correlated metal \cite{RevModPhys.83.471} and hence a good candidate for our approach. The thermopower is shown in the Supplementary Material, and displays behavior widely seen in e.g. dynamical mean field theory (DMFT) studies of strongly correlated systems \cite{pruschke, PhysRevLett.80.4775, PhysRevB.61.7996, PhysRevB.89.155101}: large values $S \sim k_B/e$ and changes in sign as a function of temperature.

At the highest temperatures $k_B T \gg U,V$ our numerical results are in excellent agreement with known expressions for the standard on-site Hubbard model \cite{PhysRevB.10.2186,PhysRevB.13.647, doi:10.1063/1.2712775}. 
We summarize these results in the Supplementary Material. The salient features are an exact $T$-linear resistivity, a temperature-independent thermopower $S$ and a Lorenz ratio $L \sim 1/T^2$. These limiting behaviors are largely independent of the interactions \cite{PhysRevLett.80.4775, Hartnoll:2014lpa, PhysRevB.95.041110, PhysRevB.94.235115}.



{\it Origin of T-linear resistivity.---} The Gaussian zero frequency peak in $\sigma(\omega)$ can be fit to
\be\label{eq:drude}
\sigma_1(\omega) = {\mathcal D} \tau \, e^{- \pi (\tau \omega)^2} \,.
\ee
Thus $\tau$ is the current relaxation or transport lifetime and ${\mathcal D}$ is the `Drude weight'. The resistivity is then $\rho = 1/({\mathcal D} \tau)$. In our incoherent regime, ${\mathcal D}$ in (\ref{eq:drude}) is best thought of as the total kinetic energy of electrons contributing to the low frequency conductance peak \cite{RevModPhys.83.471}. 

\begin{figure}[h]
\centering
\includegraphics[width=0.96\columnwidth]{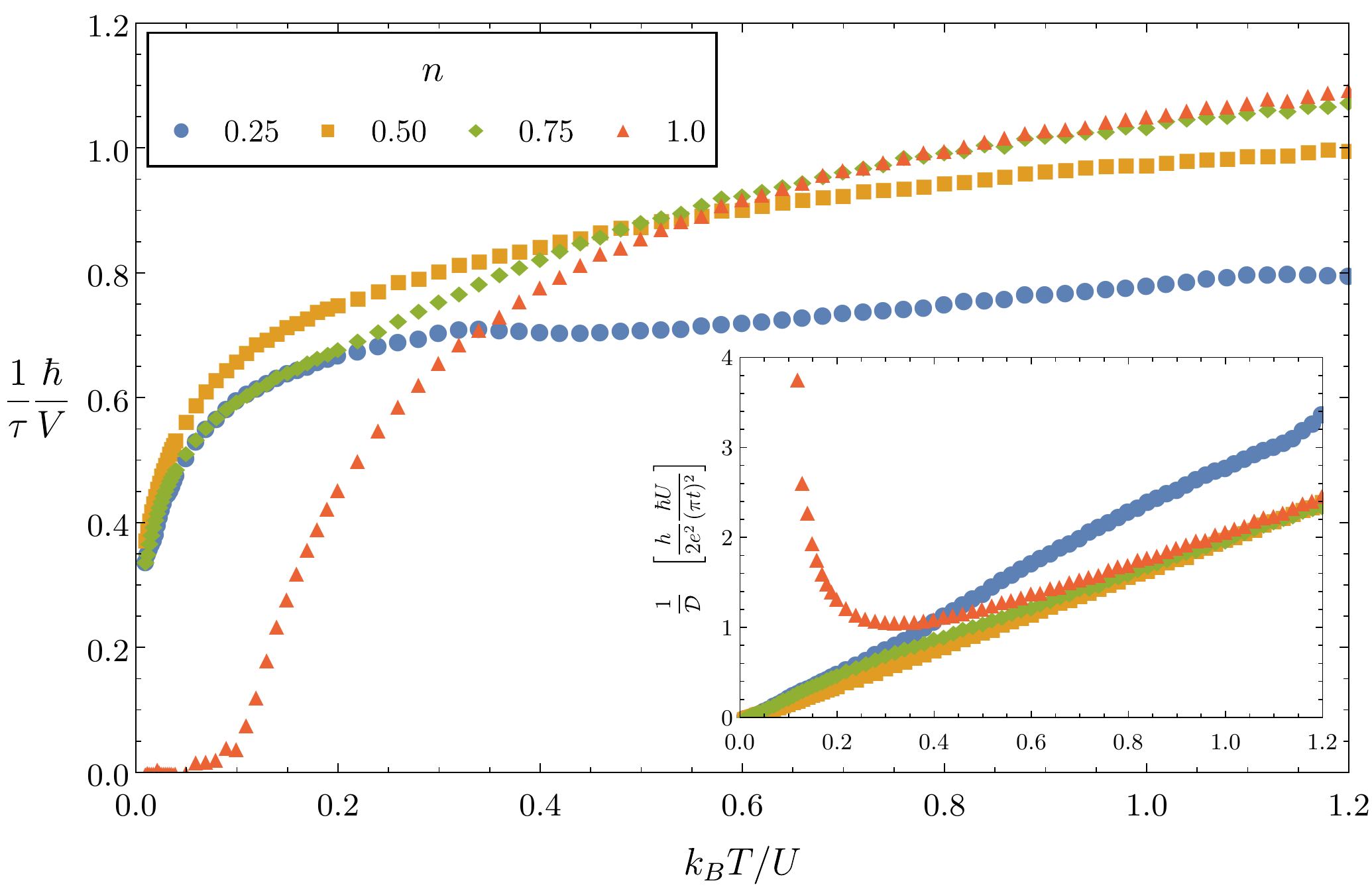}
\caption{Current relaxation rate as a function of temperature. Inset: Inverse Drude weight as of function of temperature. The statistical uncertainty in the fit to (\ref{eq:drude}) is negligible.}
\label{fig:tauandD}
\end{figure}

Figure \ref{fig:tauandD} shows the current relaxation rate $1/\tau$ as a function of temperature. The relaxation rates all saturate to a constant of order $V/\hbar$ at high $T$. The $V$ interaction is responsible for the finite transport lifetime at small $t$, whereas in the Hubbard model this lifetime must be generated nonperturbatively in $t$. Away from half filling, the relaxation rate becomes only mildly temperature-dependent below $k_B T \sim U$ and remains nonzero at the lowest temperatures we have probed \footnote{At temperatures $k_B T \ll V$, we expect charge ordering away from half filling. Indeed, working with larger values of $V \sim U$, strong features in the low temperature specific heat are seen at commensurate fillings of $n=\half, \qtr, \cdots$, at which CDW ordering occurs. The spins remain disordered (even at $n=1$) because the $U$ and $V$ interactions only depend on the charge density. This low temperature physics will change once $t$ is treated beyond perturbation theory and is furthermore unrelated to our discussion of bad metallicity, so we have not studied it in detail.}. The approximate $T$-linearity of the resistivity over this temperature range is instead controlled by the kinetic energy of the conduction electrons, which exhibits a strong temperature dependence ${\mathcal D} \sim t^2/T$, shown in the inset of figure \ref{fig:tauandD}. The decrease of ${\mathcal D}$ with increasing temperature is due to increasingly random single particle kinetic energies of both signs, that tend to cancel. We expect the low temperature divergence in ${\mathcal D}$ to be cut off below $T \sim t$, crossing over to the Fermi liquid value ${\mathcal D} \sim t$.

The total kinetic energy of all electrons can be written $K_\text{tot} \equiv \int_{-\infty}^\infty \sigma(\omega) d\omega$. The ratio ${\mathcal D}/K_\text{tot}$ therefore measures the reduction of the conductance peak kinetic energy due to interactions. Figure \ref{fig:WFTlow} shows this ratio across the phase diagram. The values of ${\mathcal D}/K_\text{tot} \sim 0.4 - 0.6$ seen in the proximity of the Mott regime are characteristic of those observed in strongly correlated metals \cite{RevModPhys.83.471}.

{\it Distinct bad metal regimes.---} Hidden under the featureless $T$-linear resistivity lies a crossover in behavior at $k_B T \sim U$. There are in fact two bad metallic regimes in the model; temperatures $k_B T \lesssim U$ are physically distinct from the infinite temperature limit. This can be seen by considering the diffusivity.

\begin{figure}[h]
\centering
\includegraphics[width=\columnwidth]{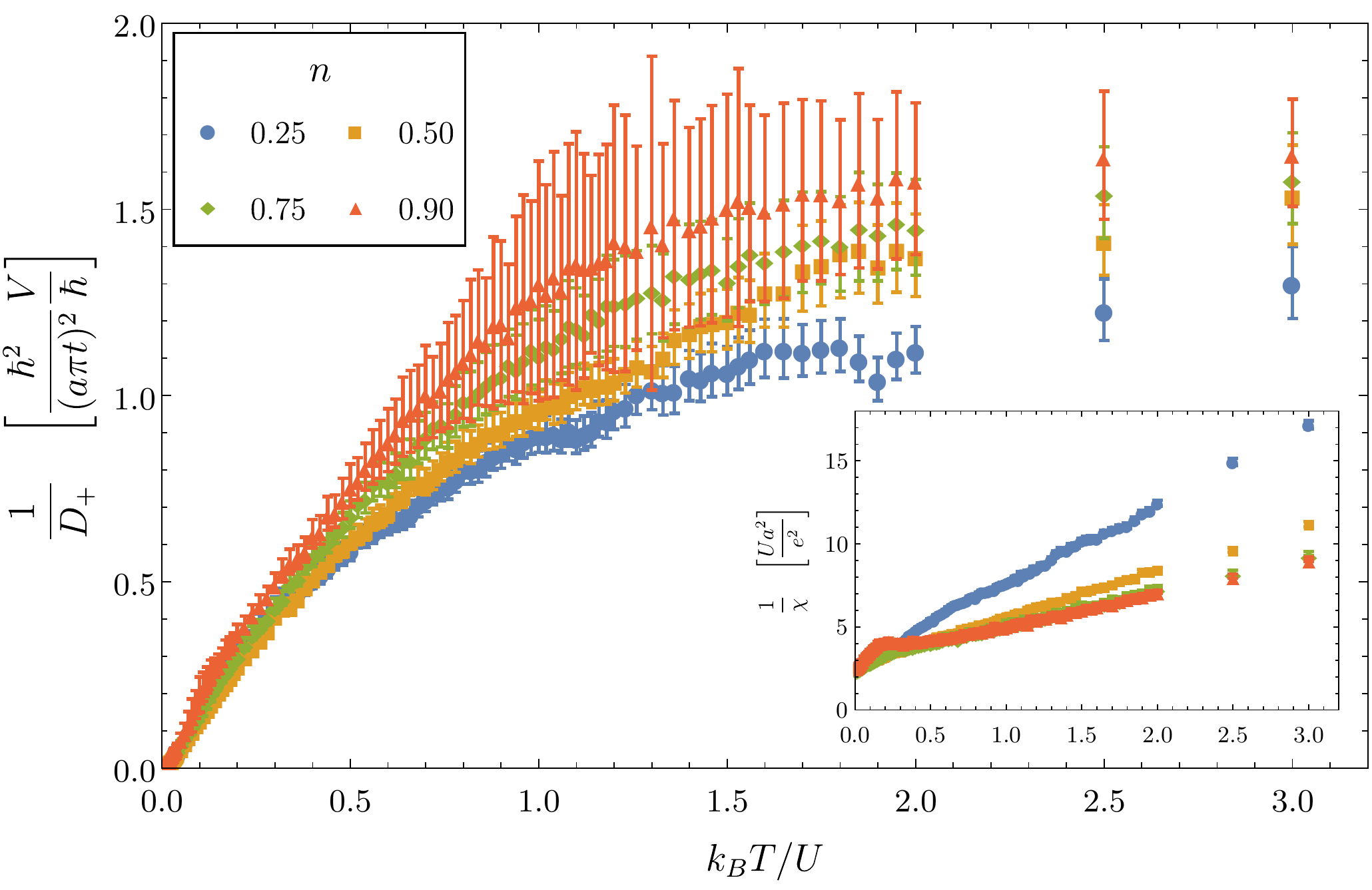}
\caption{Inverse diffusivity against temperature. The larger errors bars on the diffusivity are due to a near-cancellation in the computation of $\kappa$ and $c_n$, see Supplementary Material. Inset: Inverse susceptibility against temperature.}
\label{fig:D}
\end{figure}

In the small $t$ regime it is necessary to consider coupled charge and heat diffusion. There are three conductivities $\sigma,\alpha$ and $\kappa$ and three associated thermodynamic susceptibilities: $\chi \equiv - e^2 \, \pa^2 f/\pa \mu^2$, $\zeta \equiv - e \, \pa^2 f/\pa T \pa \mu$ and $c_{\mu} \equiv - T \, \pa^2 f/\pa T^2$, as well as  the specific heat at fixed charge $c_n \equiv c_\mu - T \zeta^2/\chi$. These determine two independent diffusivities $D_\pm$ by \cite{Hartnoll:2014lpa}: $D_+ D_- = \sigma/\chi \cdot \kappa/c_n$ and $D_+ + D_- = \sigma/\chi + \kappa/c_n + T(\zeta \sigma - \chi \alpha)^2/(c_n \chi^2 \sigma)$.
Figure \ref{fig:D} shows the diffusivity $D_+$ as a function of temperature for several fillings.
The behavior of $D_-$ is similar. The diffusivities are temperature-dependent below $k_BT \sim U$ but constant at high temperatures. We have extended the temperature range to make the saturation clearer. 

The susceptibilities also exhibit crossovers at $k_B T \sim U$. For example, the charge compressibility $\chi$ is well-described by $1/\chi = a + b \, T/U$, for doping-dependent coefficients $a$ and $b$. See inset of figure \ref{fig:D}. The nontrivial temperature dependence of the diffusivities and thermodynamic susceptibilities conspire to cancel out of the electrical resistivity, whose approximately $T$-linear behavior is featureless across $k_B T \sim U$, as found in \cite{Bakr,tom}.

The high temperature behavior of $D_\pm$ follows from the Hubbard model formulae collected in the Supplementary Material: $D_\pm = c_\pm \tau (a \pi t)^2/\hbar^2 \left[ 1 + \ocal(V/U) \right]$, with $c_+ = 2/\pi$ and $c_- = n(2-n)/\pi$. Recall that $\tau$ is temperature-independent at high temperatures.
Writing $D_\pm \lesssim \half v^2 \tau$, these expressions reveal the expected `Lieb-Robinson'-like microscopic operator growth velocity of $v \sim a \pi t/\hbar$, in the sense of \cite{Hartman:2017hhp}. At temperatures $k_B T \lesssim U$, the effective velocity $v_\text{eff}^2 \equiv 2 D_+/\tau$ becomes temperature dependent, tracking the temperature dependence of the kinetic energy ${\mathcal D}$, discussed above.

{\it Origin of bad metallic transport.---} Figure \ref{fig:Histogram}, top right, shows an interaction-induced, emergent disordered landscape of on-site potentials. The current decay rate is set by the strength of inhomogeneities in this landscape: $1/\tau \sim \Delta \epsilon \sim V/\hbar$. The separation of scales $t\ll U,V$ implies that the landscape evolves slowly, and is static on the timescale of current decay. Therefore, while momentum is microscopically relaxed by umklapp-like electronic interactions, transport is effectively controlled by local hops in an inhomogeneous potential. The usual arguments for a Mott-Ioffe-Regel bound are thus inapplicable because current is not carried by delocalized excitations with a well-defined momentum. This is the same reason that the bound does not apply to free electrons in a disordered background potential, and raises the concern that our interacting model may similarly exhibit localization.



Indeed, we noted above that the small $t$ perturbative computation of the conductivity is not strictly valid for low frequencies $\omega\in(-t,t)$. We will not exclude the possibility that a gap opens in this frequency range, analogously to how the Mott argument leads to a soft gap for strongly disordered free electrons \cite{doi:10.1080/14786436808223201}. Interactions can reduce the strength of the Mott argument due to an increased many-body phase space \cite{PhysRevB.92.104202}. Most importantly, however, even if such many-body localization does occur in our model, it is fragile and can be destroyed by coupling to physical degrees of freedom that have been omitted for simplicity in the model. As a proof of concept, we show in the Supplementary Material that coupling our model to phonons with a Debye scale $\omega_0$ and dimensionless electron-phonon coupling $g$ smears out any low frequency gap if $t \ll \sqrt{g \, \omega_0 \, k_B T} \ll U,V,k_{B}T$, while leaving our transport results intact.

{\it Discussion.---} Recent transport measurements in a cold-atomic realization of the Hubbard model with $0 \leq k_B T \lesssim U$ and $t \ll U$ show remarkable similarities with our results \cite{Bakr}. As we have found, the experiments show a nontrivial temperature dependence of the diffusivity and charge susceptibility, that cancel out to produce a close to $T$-linear resistivity. The individual temperature dependence and magnitude of these quantities are all similar to those that we have found. This suggests that, at least for temperatures $k_B T \gtrsim t$, our $V$ interaction captures the same physics as a nonperturbative treatment of $t$ in the on-site Hubbard model. Indeed, our results are also in agreement with the trends observed in Quantum Monte Carlo simulation of transport in a Hubbard model \cite{tom} over a similar temperature range, with real time transport behavior inferred from the Euclidean data \cite{Lederer4905}.

Finally, a hierarchy between the current decay rate and the single particle bandwidth also underpins an interesting recent body of work on strange and bad metals in large $N$ models \cite{Hartnoll:2016apf, PhysRevB.59.5341, PhysRevLett.119.216601, Patel:2017mjv, Chowdhury:2018sho, Berg1, Berg2} and DMFT \cite{PhysRevLett.110.086401, PhysRevLett.114.246402}. In our model this hierarchy was realized by an emergent, strongly disordered landscape. Those approaches instead effectively provide an inert `bath' into which current-carrying excitations can decay. In weakly interacting descriptions of transport, based on the Boltzmann equation, such inert scattering backgrounds have long played an important role in e.g. Bloch's formulation of the electron-phonon problem \cite{ziman} or in the Hlubina-Rice description of scattering from quantum critical spin modes \cite{PhysRevB.51.9253}. However, within strongly interacting approaches to bad metals, additional control parameters such as large $N$ or large coordination number have been necessary in order to decouple the dynamics of the bath from the current-carrying excitations.

\section*{Acknowledgements}

We have benefited from many helpful discussions with Steve Kivelson. We thank Sam Lederer for helpful comments on an earlier draft. This work of SAH and CHM has been partially supported by seed funding from SIMES. CHM is also partially supported by an NSF graduate fellowship. IE was supported by the U.S. Department of Energy, Office of Basic Energy Sciences, Division of Materials Sciences and Engineering, under Contract No. DE-AC02-76SF00515. Computational work was performed on the Sherlock cluster at Stanford University.

\newpage

\appendix

\section*{Supplementary Material}

{\it Derivation of the conductivity formulae.---} The electric and thermal current operators associated with the Hamiltonian (\ref{eq:model}) are -- see e.g. \cite{PhysRevB.67.115131} for a discussion of thermal current operators --
\begin{align}
J^{\hat \alpha} & = \displaystyle - i e \frac{ t a}{\hbar} \sum_{is} \left( c^\dagger_{i s} c^{\phantom{\dagger}}_{(i - \hat \a) s} - c^\dagger_{(i - \hat \a) s} c^{\phantom{\dagger}}_{i s} \right) \,, \label{eq:J} \\
Q^{\hat \alpha} & =  \displaystyle -i \frac{ t a}{\hbar} \sum_{is} E_{{\hat \alpha}is} \left(
c^\dagger_{i s} 
c^{\phantom{\dagger}}_{(i - \hat \a) s} 
- 
c^\dagger_{(i - \hat \a) s} c^{\phantom{\dagger}}_{i s} \right)  \,. \label{eq:Q}
\end{align}
Here $\hat \a$ is a unit lattice shift in the $x$ or $y$ direction and
\be
E_{{\hat \alpha}is} \equiv 
\frac{\epsilon_{(i - \hat \a) s} 
+ \epsilon_{i s}}{2}
- \mu \,,
\ee
with the on-site energy operator
\be\label{eq:epsilon}
\epsilon_{i s} = U n_{i\bar s} + V \sum_{j \neq i} e^{ - |\vec x_i - \vec x_j|/\ell} \, n_j \,,
\ee
where $n_{i \bar s}$ is the number of electrons at site $i$ with opposite spin to $s$. The heat current $Q$ as written in (\ref{eq:Q}) drops a $\sim t^2$ term which we ignore in our perturbative approach. As usual, the thermoelectric conductivities are obtained from the current operators using Kubo formula:
\be
\sigma=\sigma_{J^xJ^x},~~~\alpha=\frac{1}{T} \sigma_{J^xQ^x},~~~\bar{\kappa}=\frac{1}{T} \sigma_{Q^xQ^x} \,,
\ee
where
\be\label{eq:sAB}
\sigma_{AB}(\omega)
\equiv \frac{1}{\text{vol}}\int_{0}^{\infty} d\t e^{i\omega^{+}\t}
\int_0^\beta d\lambda \langle A(\t-i\lambda)B\rangle_{\beta} \,,
\ee
with $\langle \, \cdot \, \rangle_{\beta}\equiv \Tr(e^{-\beta (H-\mu N)} \, \cdot \,)/{\mathcal Z}$. 

To evaluate the Kubo formulae, the time dependence of the electrical and heat current operators is needed. The above currents are proportional to $t$. This means that to leading order in small $t$ perturbation theory, the Kubo formulae correlation functions can be evaluated in the $t=0$ theory. In particular,
the operators can be evolved setting $t=0$ in the Hamiltonian (\ref{eq:model}). Straightforward manipulations then show that the time evolution of the currents is given by making the following replacements in (\ref{eq:J}) and (\ref{eq:Q}):
\begin{align}
c^\dagger_{i s} 
c^{\phantom{\dagger}}_{(i - \hat \a) s} & \to e^{i\tau \Delta \epsilon_{i s}}
c^\dagger_{i s} 
c^{\phantom{\dagger}}_{(i - \hat \a) s} \,, \\
c^\dagger_{(i - \hat \a) s} c^{\phantom{\dagger}}_{i s} & \to e^{-i\tau \Delta \epsilon_{i s}}
c^\dagger_{(i - \hat \a) s} c^{\phantom{\dagger}}_{i s} \,.
\end{align}
The energy differences $\Delta \epsilon_{is} \equiv \epsilon_{i s} - \epsilon_{(i-\hat\alpha) s}$ are created by a single electron hopping between neighbouring sites.

We can similarly set $t=0$ in the Hamiltonian that appears in the thermal expectation value trace. This means that the trace itself is most easily performed in the occupation number basis, where the $t=0$ Hamiltonian is diagonal. Thus, thermal expectation values are computed to leading order in small $t$ utilizing a sum over occupation number configurations, $\sum_{\{n\}}$. After performing the two integrals in (\ref{eq:sAB}):
\begin{align}
\sigma(\omega)&=
\frac{t^2A e^2}{\pi}
\sum_{\{n\}} 
P_{ \{n\} }
\sum_{is}
g(\Delta \epsilon_{i s})
\, ,\\
\alpha(\omega)&=
\frac{t^2A e}{\pi T}
\sum_{\{n\}} 
P_{ \{n\} }
\sum_{is} 
E_{{\hat \alpha}is}
g(\Delta \epsilon_{i s})
,\\
\bar \kappa(\omega)&=
\frac{t^2A}{\pi T}
\sum_{\{n\}} 
P_{ \{n\} }
\sum_{is} 
E_{{\hat \alpha}is}^2
g(\Delta \epsilon_{i s})
\, .
\end{align}
In these formulae the first sum is over all configurations of occupation numbers of the lattice. These configurations are Boltzmann weighted with probabilities
\be
P_{ \{n\} } = \frac{e^{-\beta (E_{ \{n\} }-\mu N_{ \{n\} }) }}{\mathcal{Z}} \,.
\ee
The energies are $E_{\{n\}} = \frac{1}{2}
\sum_{is} n_{is} \epsilon_{is}$ and the charges are $N_{ \{n\} } = e \sum_{is} n_{is}$. $\beta \equiv 1/(k_B T)$ is the inverse temperature and $\mu$ is the chemical potential. As usual $\mathcal{Z} = \sum_{\{n\}} e^{-\beta (E_{\{n\}}-\mu N)}$. The prefactor is $A = \pi a^2/(\hbar^2 \, \text{vol})$, where $\text{vol}$ is the volume. The function $g$ is defined as
\begin{align}
g(x) & \equiv 
\frac{i}{\omega^++x} 
f(-x)
n_{is}(1-n_{(i-\hat \alpha) s}) \nonumber \\
& +
\frac{i}{\omega^+ - x} f(x)
n_{(i-\hat \alpha) s}(1-n_{is}) \,,
\end{align}
with
\be
f(x) \equiv\frac{1-e^{-\hbar \beta x}}{\hbar x} \,.
\ee
Recall that the typically measured open-circuit thermal conductivity is $\kappa\equiv \bar{\kappa} - T\alpha^2/\sigma$.

Taking the real parts:
\begin{align}
\text{Re}\,\sigma(\omega) & = t^2 A f(\omega) e^2  \sum_{\{n\}} P_{ \{n\} } \sum_{i,s} \Delta_{is}(\omega)
\,, \label{eq:sigma} \\
\text{Re}\,\a(\omega) & =  t^2  A f(\omega) \frac{e}{T} \sum_{\{n\}} P_{ \{n\} } \sum_{i,s} E_{{\hat \alpha}is} \Delta_{is}(\omega) 
\,, \\
\text{Re}\,\bar \kappa(\omega) & = t^2 A f(\omega) \frac{1}{T} \sum_{\{n\}} P_{ \{n\} } \sum_{i,s}  
E_{{\hat \alpha}is}^2 \Delta_{is}(\omega) \,. \label{eq:kappao}
\end{align}
Here $\Delta_{is}(\omega)=\Delta^+_{is}(\omega) + \Delta^-_{is}(\omega)$ are the spectral weights due to the energy differences $\Delta \epsilon_{is} \equiv \epsilon_{i s} - \epsilon_{(i-\hat\alpha) s}$:
\begin{align}
\Delta^+_{is}(\omega)  & = \delta\left(\omega + \Delta \epsilon_{is}/\hbar\right) n_{is} (1 - n_{(i-\hat \a)s}) \,, \\
\Delta^-_{is}(\omega) & =
\delta\left(\omega - \Delta \epsilon_{is}/\hbar\right) n_{(i-\hat \a)s} (1 - n_{is}) \,.
\end{align}
So long as the volume is finite, this is a sum over delta functions. These must be binned in order to approximate the smooth function that is obtained in the infinite volume limit. As noted in the main text, and as shown immediately below, the exponentially decaying interactions $H_V$ produce a continuous energy spectrum in the $t=0$ theory. The conductivities can therefore form smooth functions in frequency space to leading order in the hopping $t$.  As discussed in the main text, the above expressions for $\Delta^\pm_{is}(\omega)$ are strictly only valid for $\hbar \omega \gtrsim t$.

{\it An exponential interaction lifts the extensive degeneracy.---} The infinite range interactions $H_V$ were added in order to lift the extensive degeneracy of the $t=0$ on-site Hubbard model, thereby resolving the divergences in the conductivities. For this to work, in the limit of infinite volume, the single-particle energies must form a continuum on a finite number of bounded intervals. We will now prove that the $e^{-|\vec x|/\ell}$ potential of (\ref{eq:model}) achieves this when $\ell \geq \ell_\star \approx 1.76a$.

The strategy of the proof is ultimately to explicitly construct occupation number configurations that realize a continuum of single-particle energies at any given site. Prior to doing this, it is necessary to work the problem into a more manageable form.

The single-particle energy of a spin $s$ at position $\vec{\imath}$ is $\epsilon = Un_{\vec{\imath}\bar{s}}+\sum_{\vec{\jmath}s} V_{|\vec{\jmath}-\vec{\imath}|}n_{\vec{\jmath}s}$ with an exponentially decaying interaction $V_{r}=e^{-ra/l}$.
The first step is to replace this quantity with a different quantity that is easier to deal with. To this end, define the energy $\epsilon'$ as the interaction energy between site $\vec{\imath}$ and all sites at least $R>a$ away, so that $\epsilon' \equiv
\sum_{ |\vec{\jmath}-\vec{\imath}|\geq R,
s} V_{|\vec{\jmath}-\vec{\imath}|}n_{\vec{\jmath}s}$. 
The reason for doing this is that when $R$ is large enough we will be able to smooth out the discrete square-lattice structure that is otherwise awkward. The difference $\epsilon-\epsilon'$ has support only on a finite number of sites near $\vec{\imath}$. Therefore, as long as the values of $\epsilon'$ form a continuum for some finite choice of $R$, the energies $\epsilon$ will also form a continuum on some finite number of bounded intervals. $\epsilon'$ can be expressed more simply as $\epsilon' \equiv
\sum_{r\geq R} V_{r}N_{r}$ 
where $N_r$ counts the total number of particles a distance $r$ away from site $\vec{\imath}$. 

We can now group lattice sites at least distance $R$ away from $\vec{\imath}$ into circular shells of width $\delta$, with $\delta$ small enough so that $V_r$ can be treated as uniform within this shell. This approximation is valid as long as $R$ is large enough. The total number of particles in this shell is upper bounded by the total number of single-particle states available.
This is the area of the shell divided by $a^2$, multiplied by $2$ to account for spins, i.e. $N_{r}\leq 2\times 2\pi r \delta/a^2$. We will take $\delta=a$ for notational simplicity. We also define $M_r$ to be shifted from $N_r$ by a constant value: $M_{r}\equiv N_{r}-2\pi r \delta/a^2$, so that $M_{r}\in [-2\pi r \delta/a^2, 2\pi r \delta/a^2]$. $M_{r}$ measures the doping away from half-filling within a single shell. Then, up to an unimportant constant shift,
\begin{align}
\epsilon' = \sum_{r\geq R}V_{r}M_{r}
\equiv \sum_{s=R}^\infty \eta_{s} \alpha_{s},
\end{align}
where $\eta_{s}\equiv \text{sign}(M_{s})$ and $\alpha_{s}\equiv V_{s}|M_{s}|$. Our intention is to prove that, by choosing the signs $\eta_{s}=\pm 1$, one can force $\epsilon'$ to approach any number within a well-defined bounded interval arbitrarily closely, in the infinite volume limit. Vital to this proof is that the $\alpha_{s}\propto e^{-\frac{sa}{l}}$ approaches $0$ as $s\rightarrow\infty$. This problem is much in the same spirit as proving the Riemann series theorem.

Choosing the sign of $\eta_{s}$ corresponds to changing the number of particles $M_{s}\rightarrow-M_{s}$ in a radius-s circular shell. For example, a circular shell with filling fraction $1.4$ would transform to a filling fraction of $0.6$ under a sign change $\eta_{s}=1$ to $-1$. For each $s$, we can freely choose signs of $\eta_{s}$ while maintaining any constant filling fraction for the entire lattice by simultaneously adding or removing particles at an infinite distance away from $\vec{\imath}$, where they do not contribute to the on-site energy $\epsilon'$.

We can now explicitly construct states characterized by $\eta_{s}$ and $|M_{s}|$ which form a continuum. We choose to fix $|M_{s}|=2\pi s/a$. Then the maximal value $\epsilon'$ can take on is $A\equiv \sum_{s}\frac{2\pi s}{a}V_{s}$, which is obtained by fixing all $\eta_{s}=1$. It remains to show that the entire continuum of values in the range $[-A,A]$ can be obtained by choosing the $\eta_{s}$ appropriately. We will shortly see that, in order to do this, we need the condition  $\alpha_{n}\leq \sum_{s=n+1}^\infty \alpha_{s}$ to hold for any $n\geq R$.
That is, we need that
\begin{align}
e^{-na/l}n \leq \sum_{s=n+1}^\infty e^{-sa/l}s.
\end{align}
Setting $R\gg l$, this condition becomes $l\geq l_\star$, where $l_\star$ is defined by $e^{-a/l_\star}l_\star\approx a$ or $l_\star\approx 1.76a$. We will now show that the condition $l\geq l_\star$, where $l_\star$ serves as a minimum interaction range, is sufficient to produce a continuum in the single particle energies.

Relabelling the indices of the infinite sum to start from $1$ for convenience, we need to show that $\sum_{s=1}^\infty \eta_{s}\alpha_{s}$ can take on any value $x\in [-A,A]$. Let $0\leq x \leq A$, as any negative number can be attained by switching the sign of all $\eta_{i}$'s. Define $n_1$ so that $\eta_{1},\cdots, \eta_{n_1}=+1$ and
\begin{align}
\sum_{i=1}^{n_1-1} \eta_{i} \alpha_{i}<x,\\
\sum_{i=1}^{n_1} \eta_{i} \alpha_{i}>x.
\end{align}
This choice of $n_{1}$ must exist because $x\leq A$.  Furthermore, by construction, $|\sum_{i=1}^{n_{1}}\eta_{i}\alpha_{i}
-x|\leq \alpha_{n_1}$. Because $\alpha_{n_1}\leq \sum_{s=n_1+1}^\infty \alpha_{s}$, it must be possible to define $n_{2}>n_{1}$, with $\eta_{n_{1}+1},\cdots, \eta_{n_{2}}=-1$, so that   
\begin{align}
\sum_{i=1}^{n_2-1}\eta_{i}\alpha_i
>x,\\
\sum_{i=1}^{n_2}\eta_{i} \alpha_i
<x.
\end{align}
This procedure can be continued to define infinitely many $n_{i}$'s, with $\eta_{n_{i}+1},\cdots, \eta_{n_{i+1}}=(-1)^{i}$, so that
\begin{align}
|\sum_{i=1}^{n_N}\eta_{i}\alpha_{i}
-x|<\alpha_{n_N}.
\end{align}

The infinite sum $\sum_{i=1}^\infty \eta_{i}\alpha_{i}$ must therefore converge to $x$ because $\lim_{s\rightarrow \infty} \alpha_{s} =0$. This completes the proof that as long as $l\geq l_\star\approx 1.76a$, the single particle energies will form a continuum of values in a finite number of bounded intervals on the real line.

\begin{figure}[h]
\centering
\includegraphics[width=3.33in]{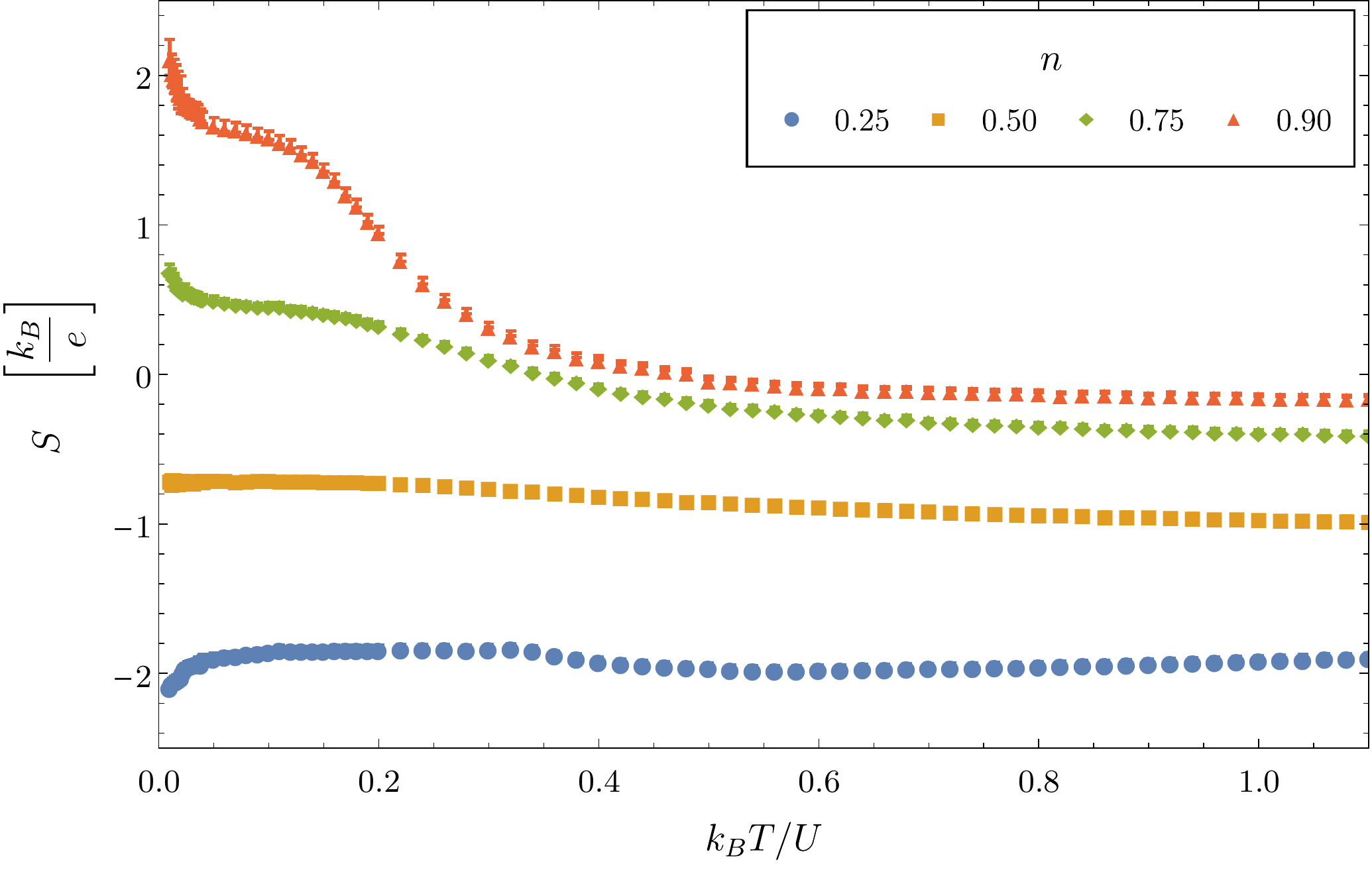}
\caption{Thermopower against temperature. Statistical Monte Carlo errors are shown. The thermopower is invariant under $n \to 2-n$ due to particle-hole symmetry of the model.}
\label{fig:WFTlow2}
\end{figure}

{\it Thermopower.---} We noted in the main text that the thermopower displays behavior characteristic of strongly correlated systems. Figure \ref{fig:WFTlow2} shows the thermopower as a function of temperature for $t \ll k_B T \lesssim U$.
The low Lorenz ratio combined with large thermopower gives a large dimensionless thermoelectric figure of merit $S^2/L$. Because we are restricted to temperatures $t \ll k_B T$, the system does not reach a Fermi liquid regime at low temperatures --- which would have led to $S$ vanishing linearly with $T$. Temperatures $k_B T \ll V$ instead lead to charge ordering, as we noted in the main text.

{\it Ultra high temperatures and the Hubbard model.---} The ultra high temperature regime $k_BT \gg U,V$ of our model is the simplest. Transport and thermodynamic quantities become very similar to those of the conventional on-site Hubbard model. The new ingredient of our model relative to the Hubbard model (or any strictly finite range extension thereof) is that all the transport coefficients are finite within small $t$ perturbation theory, due to the degeneracy-lifting $V$ interaction. Earlier works on the Hubbard model \cite{PhysRevB.10.2186, doi:10.1063/1.2712775} have regulated the low frequency divergence in transport with a `transport relaxation time' $\tau$, that is introduced by hand, and have then taken ratios of conductivities (giving, for instance, the thermopower) in which $\tau$ cancels. It is not obvious that this approach is sensible because the effective timescales that regulate thermal and electrical transport will in general be different. 
However, at high temperatures $k_B T \gg U$ the chemical potential is found to grow as $\mu \sim k_BT$. This means that the heat current $\vec Q \sim T \vec J$, from the definitions (\ref{eq:J}) and (\ref{eq:Q}). Therefore electric and thermal transport are not independent to leading order at high temperatures, and the transport lifetimes will be the same. For this reason, many of our high temperature results are the same as those for the Hubbard model.

We quickly review the Hubbard model results. In the on-site Hubbard model, the partition function can be decomposed into $\mathcal{Z}=z^N$ where $z=1+2x+x^2 e^{-\beta U}$ is the single-site partition function and $x=e^{\beta \mu}$ is the fugacity with $x=1$ at half-filling, given by
\begin{align}
x(n)
=
\frac{
-(1-n) +
\sqrt{(1-n)^2 + n (2-n) e^{-\beta U}}
}{e^{-\beta U}(2-n)}.
\end{align}

\noindent {\bf Transport:} The dc conductivities can be explicitly evaluated \cite{doi:10.1063/1.2712775} as
\begin{align}
\sigma &= \frac{e^2 t^2}{z^2} \frac{4\pi}{\hbar^2} \beta \left(x + x^3 e^{-\beta U}\right)\tau,\\
\alpha &=
\frac{e t^2}{z^2} \frac{4\pi}{\hbar^2} \beta 
\left(
\beta U x^3 e^{-\beta U} -\ln(x)(x+x^3 e^{-\beta U})
\right)\tau,\\
\bar \k &=
\frac{t^2}{z^2} \frac{4\pi}{\hbar^2} \left(
\ln(x)^2(x + x^3 e^{-\beta U}) \right. \nonumber \\
& \qquad \qquad \left. -2 \beta U \ln(x) x^3 e^{-\beta U}
+
(\beta U)^2 x^3 e^{-\beta U}
\right)\tau
,\\
\k &=
\frac{t^2 }{z^2} \frac{4\pi}{\hbar^2}(\beta U)^2
\left( \frac{x^3}{e^{\beta U}+x^2}
\right)\tau \,.
\end{align}
Recall that $\kappa\equiv \bar{\kappa} - \frac{T\alpha^2}{\sigma}$ is the open-circuit thermal conductivity. In the $t=0$ on-site Hubbard model, the transport lifetime $\tau$ is actually a $\delta(\omega)$, as emphasized in the main text. In the above formulae this has been resolved `by hand' with a single transport lifetime. In the Hubbard model this will arise due to nonperturbative in $t$ effects. In our model $\tau$ arises from the $H_V$ interaction (imagined here as a small correction to the Hubbard model).

\begin{figure}[h]
\centering
\includegraphics[width=\columnwidth]{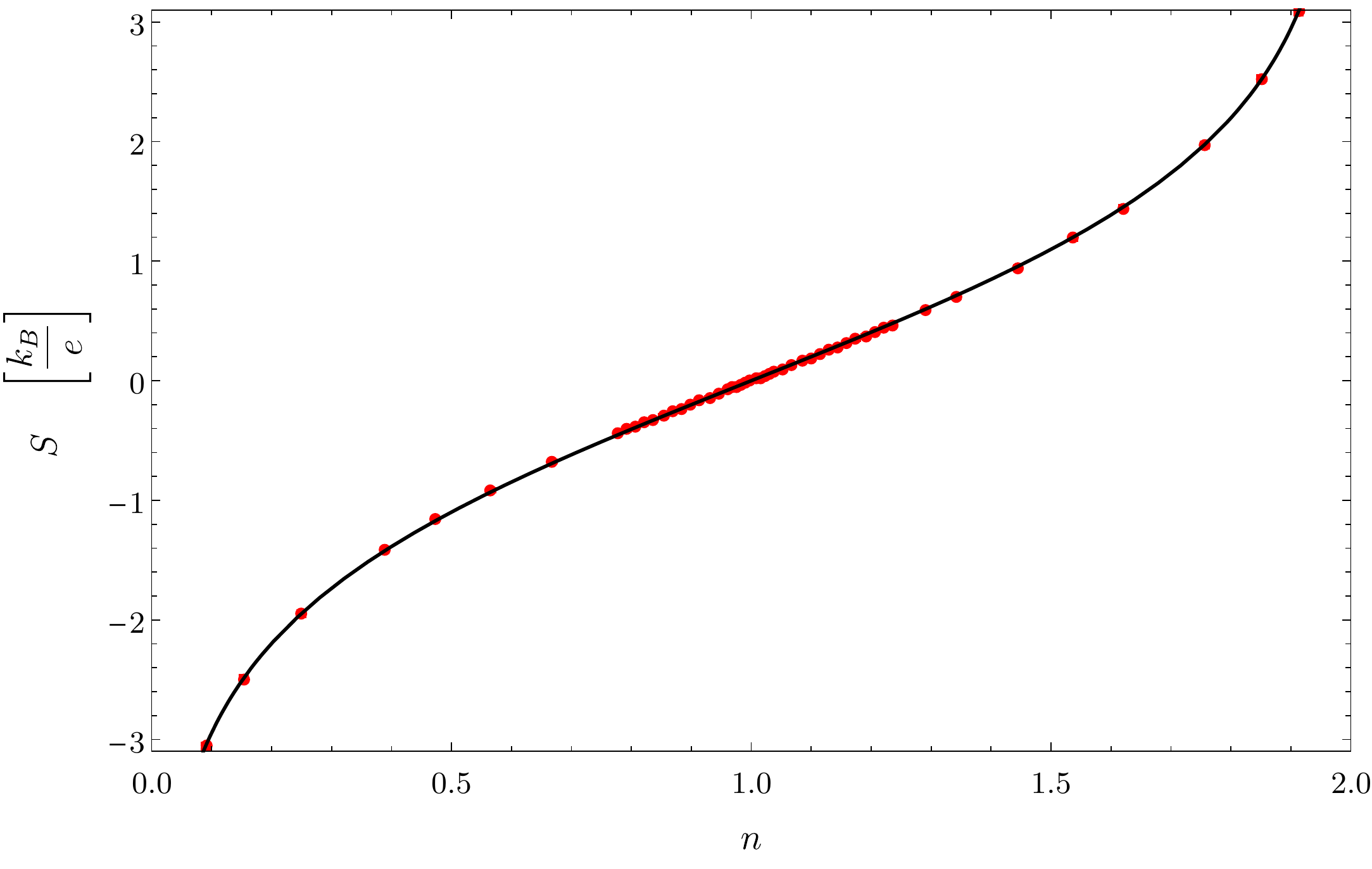}
\caption{Thermopower $S$ as a function of filling at $k_BT=10U$. The numerical data points (red) are accompanied by the on-site Hubbard model prediction derived by setting all transport lifetimes to be equal (solid black curve).}
\label{fig:SnhighT}
\end{figure}

Expanding in high temperatures, the above formulae give the expressions:
\begin{align}
\sigma &=
\frac{e^2}{h} \frac{(\pi t)^2}{k_B T} \frac{\tau}{\hbar} n (2-n)(n^2-2n+2) \,, \label{eq:ss}
\\
S & \equiv \frac{\a}{\s} = - \frac{k_B}{e} \log \frac{n}{2-n} \label{eq:SS} \,, \\
\bar L & \equiv \frac{\bar \kappa}{T \sigma} = \frac{k_B^2}{e^2} \log^2 \frac{n}{2-n} \label{eq:kk}
\,.
\end{align}
Our numerical results agree excellently with these formulae. The timescale $\tau$ is independently extracted from the width of the low frequency conductance peak, as described in the main text. The thermopower $S$ and closed circuit Lorenz ratio $\bar L$ are independent of this timescale. Figure \ref{fig:SnhighT} shows the agreement of our numerical results with the expression (\ref{eq:SS}) for $S$.

\begin{figure}[h]
\centering
\includegraphics[width=\columnwidth]{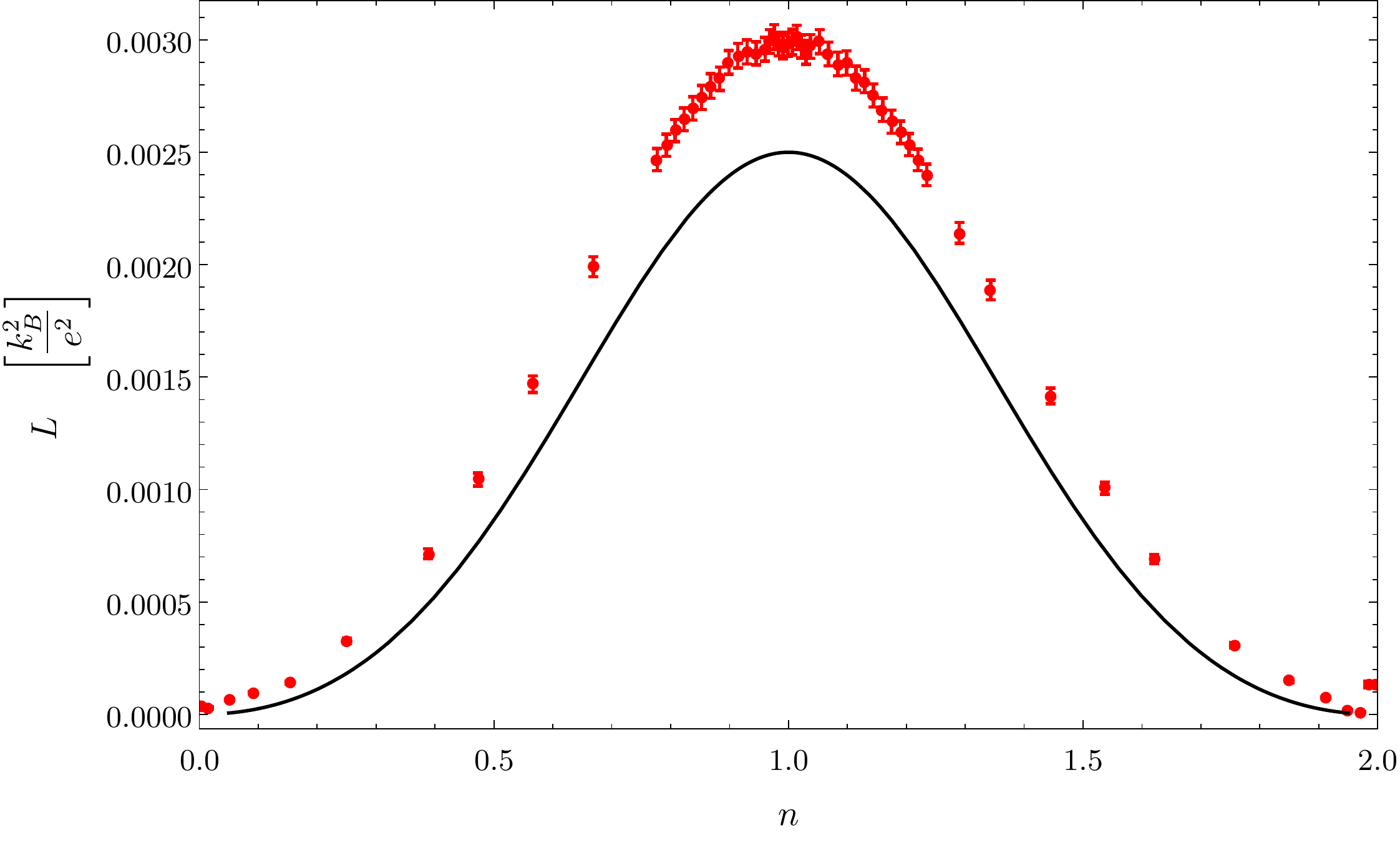}
\caption{ Lorenz ratio $L$ as a function of filling at $k_BT=10U$. Numerical data points are red, on-site Hubbard model is solid black curve.}
\label{fig:LnhighT}
\end{figure}

The closed circuit thermal conductivity $\bar \kappa$ is related to the usual open circuit conductivity by $\kappa = \bar \kappa - T \a^2/\sigma$. It is immediately seen that the two terms cancel (because $\bar L = S^2$ in (\ref{eq:kk}) and (\ref{eq:SS})), so that the thermal conductivity $\kappa$ will be additionally suppressed by factors of $U/k_BT$ in this high temperature limit. The leading order high $T$ behavior of $\k$ is therefore explicitly sensitive to the precise interactions in the model, in addition to any dependence on the relaxation lifetime. The high temperature thermal conductivity therefore only agrees with the Hubbard model up to corrections of order $V/U$. In particular the high temperature Lorenz ratio is given by
\begin{align}
L=\frac{\k}{\s T}
= \frac{k_B^2}{e^2} \frac{U^2}{(k_B T)^2} \frac{n^2(2-n)^2}{4(2-2n +n^2)^2}\Big( 1 + \ocal(V/U) \Big) \,. \nonumber
\end{align}
Figure \ref{fig:LnhighT} shows that the numerically obtained high temperature Lorenz ratio indeed agrees with the Hubbard model up to order $\mathcal{O}(\frac{V}{U})$.

\begin{figure}[h]
\centering
\includegraphics[width=\columnwidth]{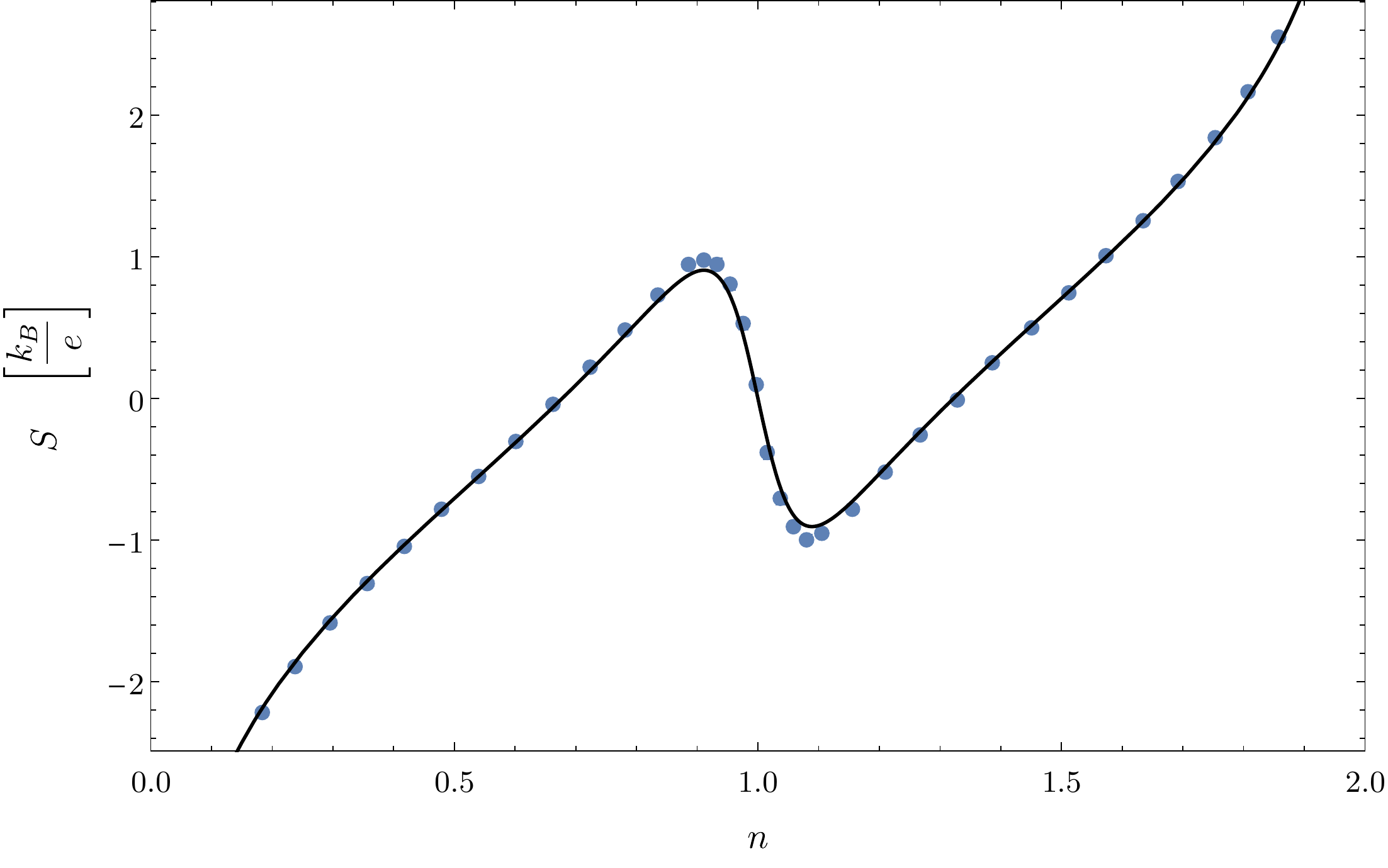}\\ \includegraphics[width=\columnwidth]{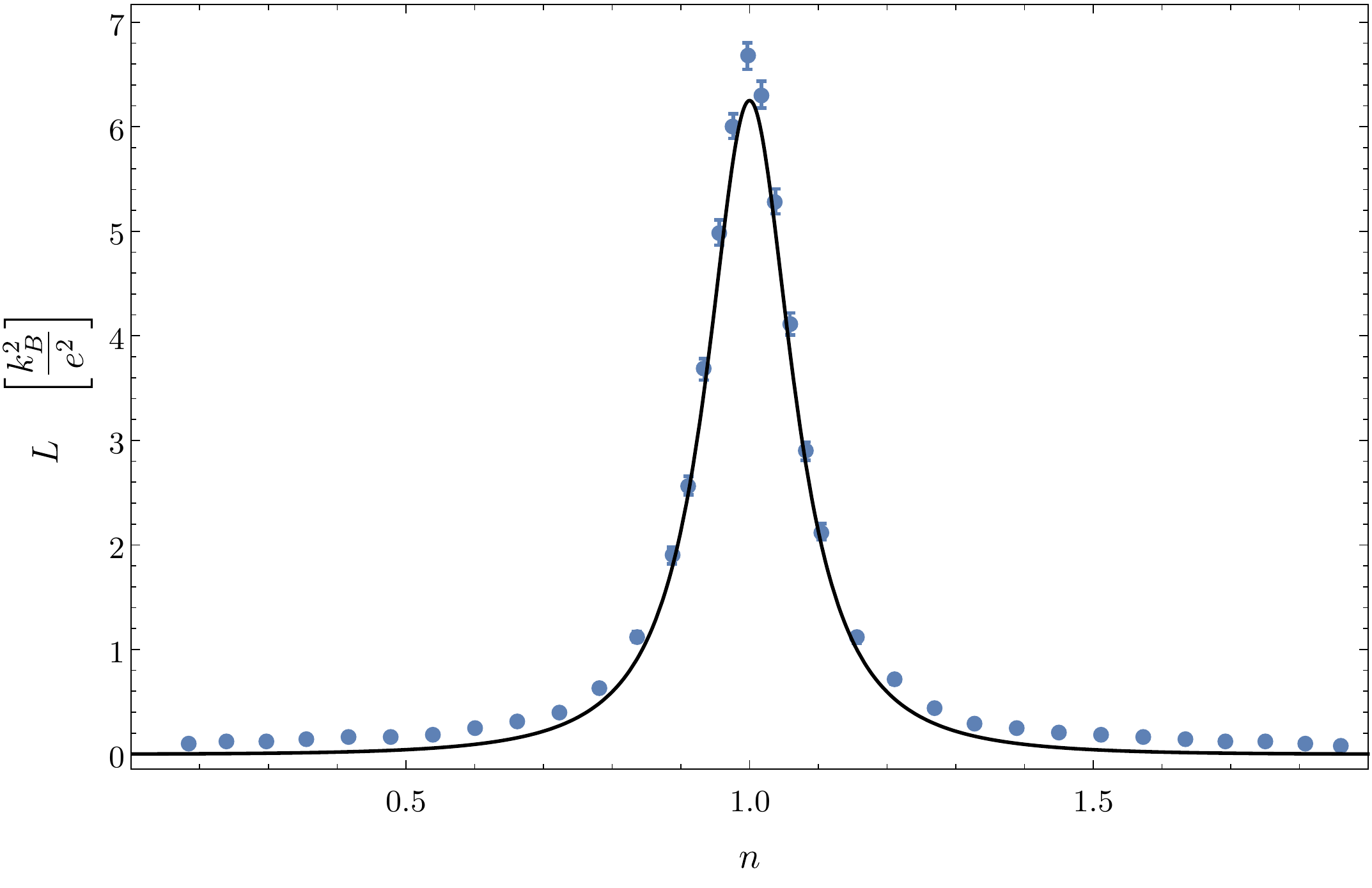}
\caption{As in figures \ref{fig:SnhighT} and \ref{fig:LnhighT}, the thermopower $S$ (top) and Lorenz ratio $L$ (bottom) as a function of filling, now at a low temperature of $k_B T=0.2U$. Numerical data points are in blue, the on-site Hubbard model predictions are shown as solid black curves.}
\label{fig:SnlowT}
\end{figure}

The Hubbard model formulae remain a good approximation for the ratios of conductivities $S$ and $L$ down to temperatures $k_BT \sim V$. As illustrated in figure \ref{fig:SnlowT}, these quantities agree up to $\mathcal{O}(\frac{V}{k_BT})$ corrections.
We note, as a passing curiosity, that the $n$ dependence of the Lorenz ratio in figure \ref{fig:SnlowT} is remarkably similar to that observed recently in a strongly correlated regime of graphene \cite{Crossno1058}, with $n=1$ playing the role of particle-hole symmetric Dirac point in graphene.

\noindent {\bf Thermodynamics:} Thermodynamic quantities can similarly be obtained exactly in the $t=0$ on-site Hubbard model. As in the main text we define
\begin{align}
\chi \equiv - e^2
\frac{\partial^2 f}{\partial \mu^2},\;
\zeta \equiv 
-e \frac{\partial^2f}{\partial T \partial \mu},\;
c_{\mu} \equiv -T \frac{\partial^2 f}{\partial T^2},\;
\end{align}
and
\begin{align}
c_{n}\equiv c_{\mu}-\frac{T \zeta^2}{\chi} \,.
\end{align}
with $f\equiv -\frac{k_B T}{\text{vol}} \ln \mathcal{Z}$ defined as the free energy density. In the on-site Hubbard model, the thermodynamic quantities are particularly simple to compute because $f=-\frac{k_{B}T}{a^2}\ln(z)$ with $z=1+2x + x^2e^{-\beta U}$. Using the identities $\partial_{\mu}x = \beta x$ and $\partial_{T}=-k_{B}\beta^2 \partial_{\beta}$, one finds that
\begin{align}
\chi 
&=
\frac{e^2}{a^2}
\frac{2\beta x}{z^2}
\left(1+
2x e^{-\beta U} + x^2 e^{-\beta U}\right),\\
\zeta
&=
\frac{ek_{B}}{a^2}
\frac{2\beta x e^{-\beta U}}{z^2}
\left(\beta Ux(1+x) \right. \nonumber \\
& \left.  \qquad \qquad \qquad - \ln(x)(e^{\beta U}+2x+x^2)\right),\\
c_{\mu}
&=\frac{k_{B}^2}{a^2}
\frac{xe^{-\beta U}}{z^2} \left(
(\beta U)^2 x (1+2x) - 4(\beta U)\ln(x)x(1+x) \right. \nonumber \\
& \left. \qquad \qquad \qquad  +
2\ln(x)^2(e^{\beta U}+2x+x^2)\right),\\
c_{n}
&=\frac{k_{B}^2}{a^2}
\frac{(\beta U)^2}{z} \frac{x^2}{x^2+2x+e^{\beta U}} \,.
\end{align}

As with the transport observables above, to leading order at high temperature our numerical results for the thermodynamic susceptibilities fit excellently to the high temperature expansions of the Hubbard model expressions:
\begin{align}
\chi &  = \frac{1}{k_B T} \frac{e^2}{a^2} \frac{n(2-n)}{2} \,, \\
\frac{\zeta}{\chi} &  =  - \frac{k_B}{e} \log \frac{n}{2-n} = S \,,\\
\frac{c_{\mu}}{T \chi}&  =  \frac{k_B^2}{e^2} \log^2 \frac{n}{2-n} = \bar L \,.
\end{align}
The specific heat at fixed density $c_n = c_\mu - T \zeta^2/\chi$ is suppressed in this high temperature limit, due to the same cancellation that occurred in the thermal conductivity $\kappa$ above.
\begin{figure}[h]
\centering
\includegraphics[width=0.5\columnwidth]{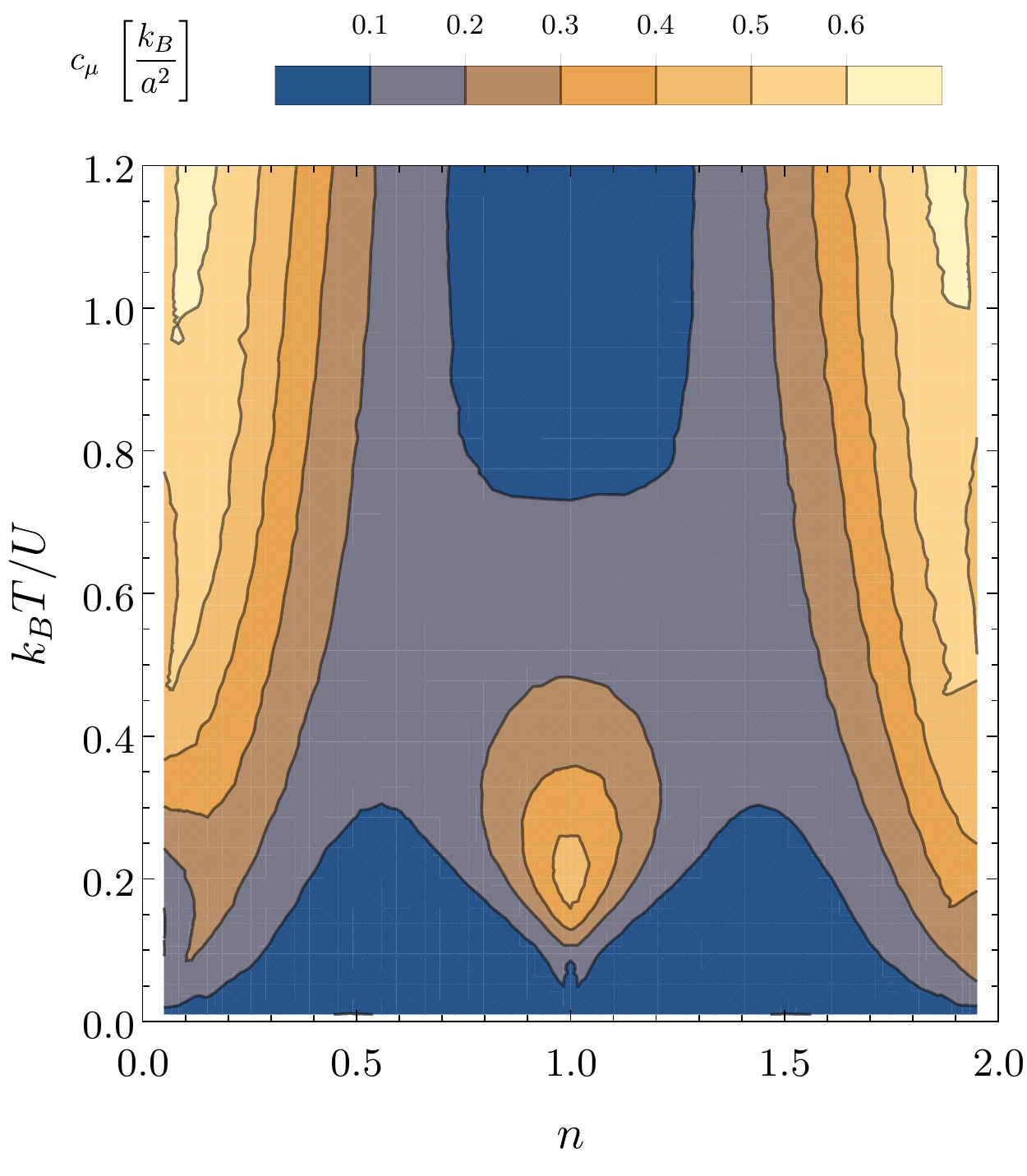}\includegraphics[width=0.5\columnwidth]{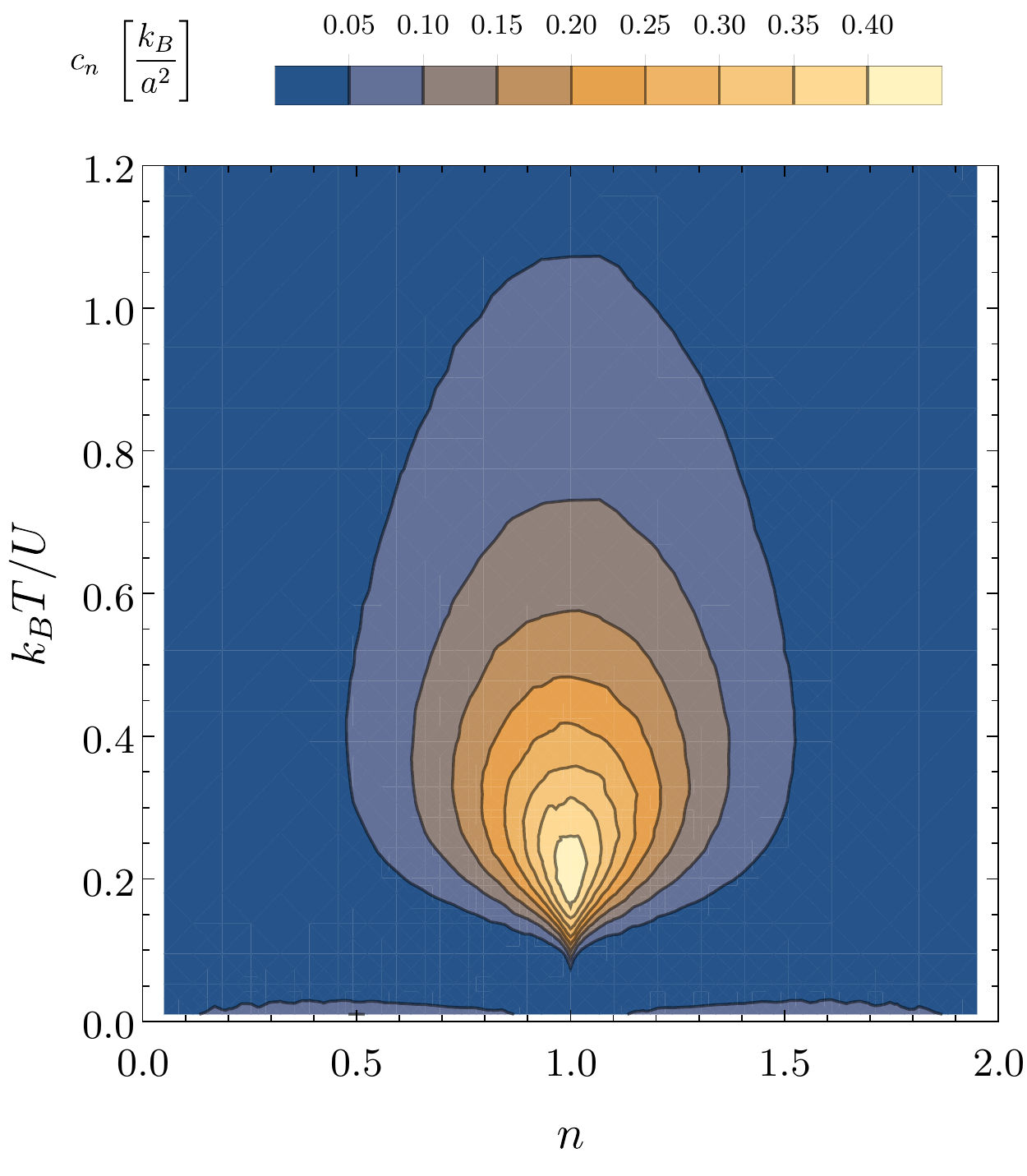}
\caption{$c_\mu$ (left) and $c_n$ (right) as a function of temperature and filling.}
\label{fig:candc}
\end{figure}

\begin{figure}[h]
\centering
\includegraphics[width=\columnwidth]{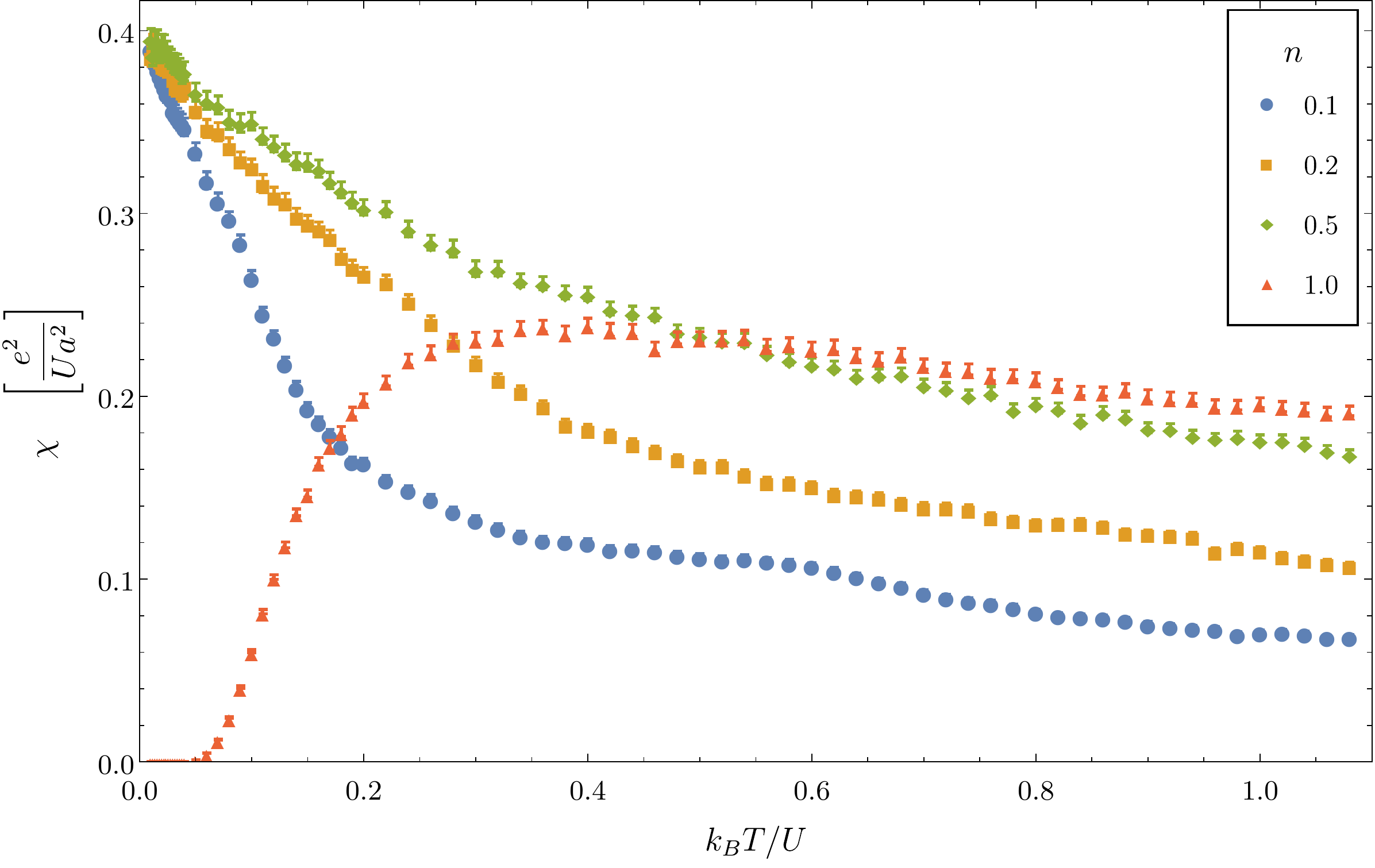}
\caption{The charge compressibility $\chi$ against temperature.}
\label{fig:Chi}
\end{figure}

In contrast to the high temperature results given above, the low temperature behavior of thermodynamic quantities is very different in our model and in the $t=0$ Hubbard model. This is because of the extensive degeneracy of the $t=0$ Hubbard model. Figure \ref{fig:candc} shows the specific heats $c_\mu$ and $c_n$ across the intermediate temperature phase diagram for our model. Figure \ref{fig:Chi} shows the susceptibility (the inverse susceptibility was already shown in the inset of figure \ref{fig:D} in the main text). The gapped Mott regime at $n=1$ is clearly visible in this figure. 

{\it Details of Monte Carlo simulation.---} We work perturbatively in the hopping parameter $t$, such that the Boltzmann factor can be approximated as $e^{-\beta H}\approx e^{-\beta (H_U+H_V)}$ as long as $\beta t\ll 1$. The interaction Hamiltonian $H_U+H_V$ consists only of occupation numbers, so that classical Monte Carlo simulation suffices to produce the thermal ensemble for the Hamiltonian. 

The Monte Carlo simulations are run across a fine mesh of points in the $(\mu,k_BT)$ plane, and the metropolis algorithm is run 15000 times for each $(\mu,k_BT)$ point. At each fixed chemical potential, the simulations are run sequentially from the highest to lowest temperatures, adiabatically cooling the samples to ensure that the configurations reach a true equilibrium. The filling fraction of each $(\mu ,k_BT)$ point is then derived from the expectation value $\langle N \rangle$ in the thermal ensemble.

\begin{figure}[h]
\centering
\includegraphics[width=0.5\columnwidth]{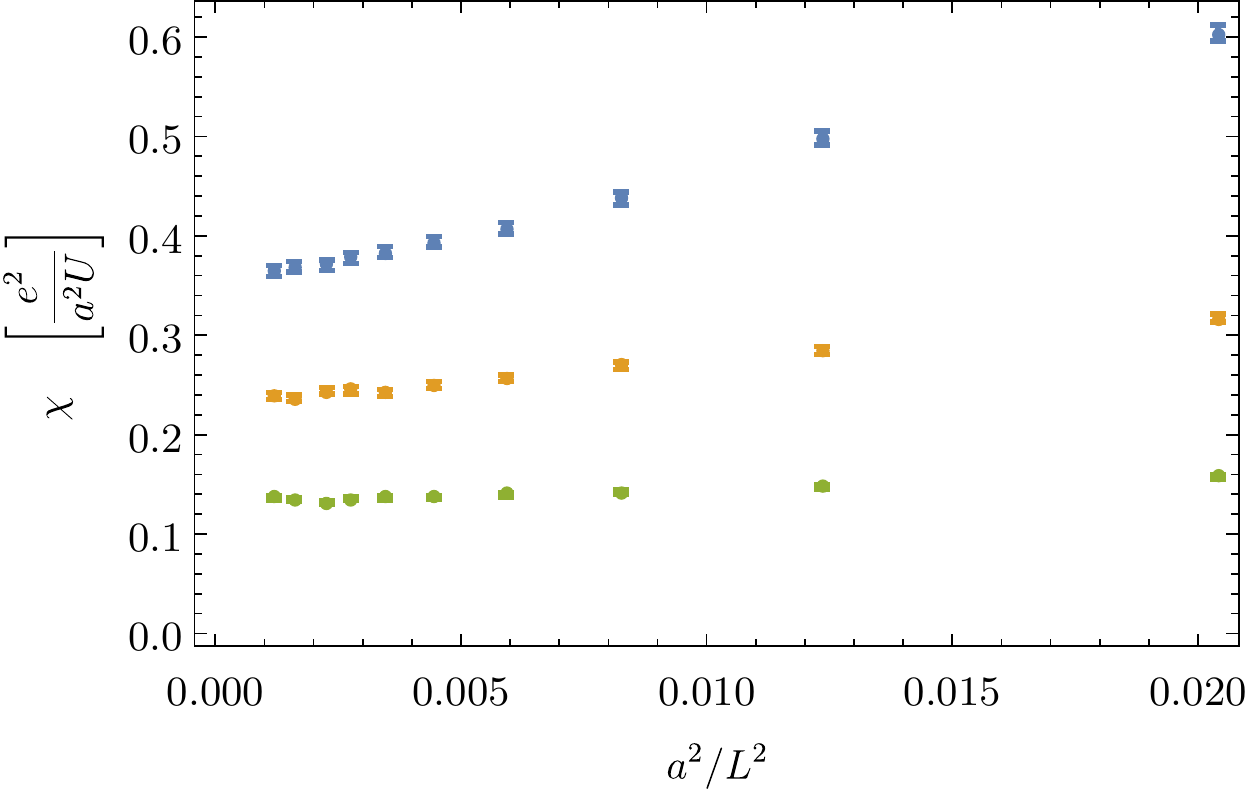}\includegraphics[width=0.5\columnwidth]{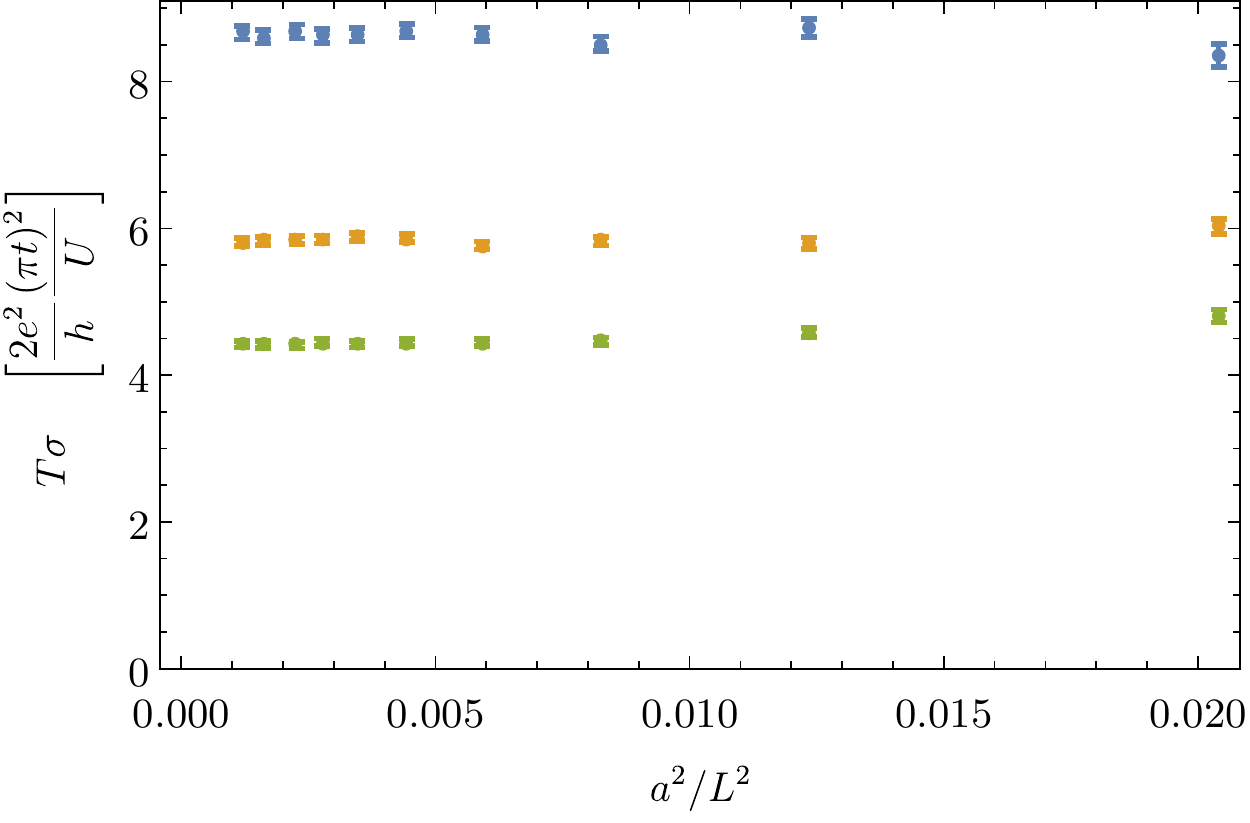}
\caption{Left: Finite size scaling of charge compressibility $\chi$. Right: Finite size scaling of DC conductivity $\sigma$ (rescaled by temperature $k_BT$). Both panels have density $n=0.7$ and are shown for three temperatures $k_BT/U = 0.05$ (blue), $0.5$ (orange), $2.0$ (green). Error bars are shown, but they are generally smaller than the symbol size.}
\label{fig:finite_size_scaling}
\end{figure}

Figure \ref{fig:finite_size_scaling} provides representative examples of the finite size effects in our results. For both the thermodynamic quantity $\chi$ and the transport quantity $\sigma$, it is immmediately evident that our results converge to the $L\rightarrow\infty$ limit very rapidly. All data shown in the main text derives from simulations for the largest system size with a side length of $L=29a$.

\newpage

{\it Smearing out many-body localization with an electron-phonon coupling.---} As discussed in the main text, we have ignored higher-order processes in $t$ which could potentially result in electron localization, introducing a (hard of soft) gap in the optical conductivity. Due to the hierarchy between the scales $t$ and $U,V$ it is straightforward for a coupling to additional degrees of freedom to prevent many-body localization. Here we show how coupling the electron-only model (\ref{eq:model}) to phonons serves this purpose, `smearing' out any potential insulating gap while otherwise not qualitatively changing the optical conductivity. This smearing energy $E_\text{smear}$ is the characteristic energy transition due to the creation or annihilation of phonons. We will require that
\be{\label{eq:smearcondition}}
E_\text{gap}\ll E_\text{smear}\ll V,U \,,
\ee
in order to smear away sharp features like the insulating gap without qualitatively affecting the rest of the optical conductivity.

For simplicity we use Einstein phonons with energy $\hbar \omega_0$, though the conclusions are easily extended for nontrival phonon dispersons. We will assume on physical grounds that $\hbar \omega_0\ll t$. The full Hamiltonian is now $H=H_\text{e} + H_\text{eph}+ H_\text{ph}$, where $H_\text{e}=H_t+H_U+H_V$ is the original electron-only Hamiltonian (\ref{eq:model}), and
\begin{align}
H_\text{eph} &= \alpha \sum_{i} n_{i} x_i \,, \\
H_\text{ph} & =\sum_{i}\frac{p_i^2}{2M}+\frac{Kx_i^2}{2} \,.
\end{align}
Here $\alpha$ is an electron-phonon coupling constant, $K$ is the spring constant, $M$ is the nucleus mass, and $x_{i}$ and $p_{i}$ are the position and momentum of the nucleus at site $i$. The well-known unitary transformation \cite{mahan2013many}
\begin{align}
\mathcal{U} = e^{i\alpha \sum_{i} p_i n_i/K}, 
\end{align}
leads to the Hamiltonian
\begin{align}
H'=
H'_{t}+ H_{U'} + H_{V}+ H_\text{ph} -\frac{\alpha^2}{K} N,
\end{align}
where $A'\equiv \mathcal{U}^\dagger A\, \mathcal{U}$, the `renormalized' Hubbard coupling $U' = U - \frac{\alpha^2}{K}$ and the density $N = \sum_{i} 
n_{i}$. The term $H_{t}'$ still couples electrons and phonons. We will assume that this coupling doesn't result in phonon localization, so that it can be treated as a (negligible) perturbation of the phonon Hamiltonian at small $t$. The electrons and phonons can then be treated as decoupled. For the electron system this term results in an effective renormalization of the hopping term. 


Using the identities  $\text{Tr} A =\text{Tr} A' $ and $(AB)'=A'B'$, the conductivity becomes
\begin{align}
\sigma_{JJ}(\omega) = 
\int_0^\beta & d\lambda  \int_{0}^\infty d\t e^{i\omega^+\t}\frac{1}{\mathcal{Z}} \times \\ & \text{Tr}(e^{-\beta (H'-\mu N)}
e^{i H'(\t-i\lambda)}
J' e^{-iH'(\t-i\lambda)}
J') \,. \nonumber
\end{align}
By using the identity $c'_{i}=X_{i} c_{i}$ with $X_{i}=e^{i\alpha p_{i}/K}$ and the observation that electrons and phonons are decoupled in $H'$, we re-express the trace over the full Hilbert space as an electron-space trace $F_{e}(\tau-i\lambda)$ and a phonon-space trace $F_{ph}(\tau-i\lambda)$:
\begin{align}
\sigma_{JJ}(\omega) = \int_0^\beta d\lambda \int_{0}^\infty d\t e^{i\omega^+\t}
F_\text{e}(\tau-i\lambda) F_\text{ph}(\tau-i\lambda),
\end{align}
Here, setting $F_{ph}(\tau-i\lambda)$ equal to $1$ would recover the electron-only optical conductivity, except with the renormalized $U'$ and hopping term. The function $F_\text{e}(\tau-i\lambda)$ Fourier transforms into a sum of delta functions for each transition energy $\Delta\epsilon_{is}$. $F_\text{ph}$ accounts for creation or annihilation events of some number $l$ of phonons, with energy $\pm l\omega_0$. Thus, the addition of phonons smears a delta function at $\Delta \epsilon_{is}$ into multiple delta functions at $\Delta \epsilon_{is} \pm l \omega_0$. In detail, following the methods of \cite{mahan2013many}, we find that
\begin{align}
F_{ph}(\tau-i\lambda)
\approx 
e^{-z}\sum_{l=-\infty}^\infty I_{l}(z) e^{il\theta},
\end{align}
where $\approx$ utilizes, for convenience, the limit $k_{B}T\gg \omega_0$, $z=2g\sqrt{n_{b}(\omega_0)(n_{b}(\omega_0)+1)}\approx 2gT/\omega_0$, $g$ is the dimensionless electron-phonon coupling $g=\alpha^2/(K\omega_0)$, $I_{l}$ is the $l$th modified Bessel function and $\theta=\omega_0(t-i\lambda+i\frac{\beta}{2})$. Each harmonic $e^{il\theta}$  corresponds to the creation or annihilation of 
$l$ phonons. The likelihood of this event is weighted by  $I_{l}(z)e^{-\beta l\omega_0/2}$. We will take $E_\text{smear}\ll k_{B}T$ so that the temperature dependence of the exponential is negligible. $I_{l}(z)$
falls off with $l$ and has a typical width as a function of $l$ of $\sim \sqrt{z}$. Therefore the phonon smearing energy is
\begin{align}
E_\text{smear}\sim \sqrt{z}\omega_0\sim \sqrt{gT\omega_0},
\end{align}

Allowing the insulating gap to take its maximal value of $E_\text{gap}\sim t$,
the requirement (\ref{eq:smearcondition}) together with the condition $E_\text{smear}\ll k_{B}T$,  imply that
\begin{align}\label{smearcondition2}
t\frac{t}{k_{B}T}\ll g\omega_0 \ll V \frac{V}{k_{B}T},k_{B}T.
\end{align}
is sufficient to smear away an insulating gap without changing the functional form of $\sigma(\omega)$. The overall amplitude and temperature dependence of $\sigma(\omega)$ are entirely unchanged by the phonons. The lower and upper bounds of (\ref{smearcondition2}) are, respectively, very small and large energy scales because $t\ll  k_{B}T,U,V$; this implies that the stated conditions are easily satisfied without fine-tuning.

\end{document}